\documentclass[aps,prd,notitlepage,nofootinbib,preprintnumbers,superscriptaddress,10pt]{revtex4-2}
\usepackage{here}
\usepackage{graphicx}
\usepackage{amsmath,amsthm,amssymb}
\usepackage{bm}
\usepackage{color}
\usepackage{multirow}

\newcommand{\be}{\begin{eqnarray}}
\newcommand{\ee}{\end{eqnarray}}

\newcommand{\bem}{\begin{bmatrix}}
\newcommand{\eem}{\end{bmatrix}}

\usepackage{amsfonts}
\usepackage{dcolumn}
\usepackage{hyperref}
\allowdisplaybreaks[1]
\usepackage{stackengine}

\allowdisplaybreaks
\begin{document}
\preprint{YITP-24-80, IPMU24-0029}

\title{
Even-parity black hole perturbations in Minimal Theory of Bigravity
}

\author{Masato Minamitsuji}
\affiliation{Faculty of Health Sciences, Butsuryo College of Osaka, Sakai, 593-8328, Osaka, Japan}
\affiliation{Centro de Astrof\'{\i}sica e Gravita\c c\~ao  - CENTRA, Departamento de F\'{\i}sica, Instituto Superior T\'ecnico - IST, Universidade de Lisboa - UL, Av. Rovisco Pais 1, 1049-001 Lisboa, Portugal}

\author{Shinji Mukohyama}
\affiliation{Center for Gravitational Physics and Quantum Information, Yukawa Institute for Theoretical Physics, Kyoto University, 606-8502, Kyoto, Japan}
\affiliation{Kavli Institute for the Physics and Mathematics of the Universe (WPI), The University of Tokyo Institutes for Advanced Study, The University of Tokyo, Kashiwa, Chiba 277-8583, Japan}

\author{Michele Oliosi}
\affiliation{PVsyst SA, Rte de la Maison-Carrée 30, 1242 Satigny, Switzerland}

\begin{abstract}
We study even-parity black hole perturbations in Minimal Theory of Bigravity (MTBG). We consider the Schwarzschild solution written in the spatially-flat coordinates in the self-accelerating branch as the background solution. We clarify the gauge transformations for the $\ell=0$, $1$, and $\geq2$ modes with $\ell$ being the angular multipole moments under the joint foliation-preserving diffeomorphism transformation. Requiring that the asymptotic regions in the physical and fiducial sectors share the same Minkowski vacua, the solution to the $\ell=0$ perturbations can be absorbed by a redefinition of the Schwarzschild background. In order to analyze the $\ell=1$ and $\geq 2$ modes, for simplicity we focus on the effectively massless case, where the constant parameter measuring the ratio of the proper times between the two sectors is set to unity and the effective mass terms in the equations of motion vanish. We also find that as a particular solution all the $\ell=1$ perturbations vanish by imposing their regularity at spatial infinity. For each of the $\ell\geq 2$ modes,  in the effectively massless case, we highlight the existence of the expected two propagating modes and four instantaneous modes.
\end{abstract}
\maketitle

\section{Introduction}
\label{sec1}

Ghost-free massive gravity and bigravity theories are promising candidates to elucidate the origin of the present day's cosmic acceleration.  
The first model of ghost-free massive gravity was formulated within a linearized theory by Fierz and Pauli~\cite{Fierz:1939ix}. 
While naive nonlinear extensions of the Fierz-Pauli theory have not been successful because of the appearance of the Boulware–Deser (BD) ghost \cite{Boulware:1972yco}, the first model of massive gravity free from the BD ghost at the fully nonlinear level was provided by de Rham, Gabadadze, and Tolley (dRGT)~\cite{deRham:2010kj}.  
dRGT model was then extended to a bigravity theory by Hassan and Rosen (HR) \cite{Hassan:2011zd}, by promoting the second fiducial metric to be another dynamical field~\footnote{HR bigravity still suffers from the BD ghost if matter is coupled to both the physical and fiducial metrics \cite{Yamashita:2014fga,deRham:2014fha,Gumrukcuoglu:2015nua}.}.

HR bigravity was extended to Minimal Theory of Bigravity (MTBG) \cite{DeFelice:2020ecp}, where the four-dimensional spacetime diffeomorphism invariance is broken down to the three-dimensional spatial diffeomorphism invariance and time-reparameterization invariance.
While the four-dimensional diffeomorphism invariance is explicitly broken, the absence of problematic scalar and vector degrees of freedom (DOFs) makes it easy for the theory to be consistent with experimental and observational tests. 
MTBG shares the same background cosmological dynamics with HR bigravity, but the number of propagating DOFs is down to four, two massless tensorial DOFs and the other two tensorial DOFs that are massive.
The absence of the extra scalar and vector DOFs in MTBG implies the absence of ghost or gradient instabilities associated with them \cite{Garcia-Saenz:2021uyv,Silva:2021jya,Demirboga:2021nrc}.
In the normal branch, deviations from GR in the dynamics of both background and the scalar sector could be already observed.
The absence of extra DOFs also allows for a new production scenario of spin-2 dark matter based on the transition from an anisotropic fixed point solution to an isotropic one~\cite{Manita:2022tkl}.

MTBG possesses constraints by which the unwanted modes can be removed nonlinearly from the theory. A consequence of the presence of these constraints is the appearance of instantaneous (or shadowy) modes~\cite{DeFelice:2018ewo,Minamitsuji:2023lvi}, which are described by elliptic equations on a three-dimensional hypersurface.
Such instantaneous modes appear not only in MTBG but also in other theories, for instance, higher-order scalar-tensor theories where the degeneracy conditions are met only in the unitary gauge, known as U-DHOST theories~\cite{DeFelice:2020eju,DeFelice:2021hps}.
They satisfy elliptic equations for which some appropriate boundary conditions need to be imposed. 

Ref.~\cite{Minamitsuji:2022vfv}  investigated static and spherically symmetric solutions in the self-accelerating and normal branches of MTBG.
It was shown that a pair of Schwarzschild-de~Sitter spacetimes with different cosmological constants and black hole (BH) masses written in the spatially-flat (Gullstrand–Painlev\'e (GP)) coordinates is a solution in the self-accelerating branch of MTBG.  
On the other hand, in the normal branch, while the spatially-flat coordinates of the paired Schwarzschild-de~Sitter metrics cannot be solutions, those written in the coordinates with a constant trace of the extrinsic curvature tensors on the constant time hypersurfaces \cite{DeFelice:2020eju,DeFelice:2020cpt,DeFelice:2020onz,DeFelice:2022uxv} could be solutions, provided that the two metrics are parallel.

Ref.~\cite{Minamitsuji:2023lvi} investigated spherically symmetric gravitational collapse of pressure-less dust in the self-accelerating branch of MTBG.
While the interior region of a collapsing solution is described by a Friedmann-Lema\^itre-Robertson-Walker (FLRW) universe, the exterior region is described by a Schwarzschild spacetime with specific time slicings.
The collapsing solution with the spatially-flat slicings has been obtained under certain tuning  of the initial conditions. 
In the spatially-closed case corresponding to an extension of the Oppenheimer-Snyder model \cite{Oppenheimer:1939ue,Kanai:2010ae,Blau}, gravitational collapse happens in the physical and fiducial sectors in the same manner as it would in two independent copies of GR under a certain tuning of the matter energy densities and Schwarzschild radii between the two sectors.

Ref.~\cite{Minamitsuji:2023lvi} also studied odd-parity perturbations of the Schwarzschild-de~Sitter solutions written in the spatially-flat coordinates in the self-accelerating branch of MTBG.
For the modes $\ell\geq 2$, where $\ell$ represents the angular multipole moment, there are four physical modes, where two of them are dynamical and the remaining two are instantaneous.
Beside the case in which the ratio of the lapse functions in the physical and fiducial sectors are equal to unity, the two dynamical modes are coupled to each other and sourced by the two instantaneous modes.
For the dipolar mode $\ell=1$, the two copies of the slow-rotation limit of the Kerr-de~Sitter metrics, in general, cannot be a solution in MTBG, indicating deviation from GR for rotating black holes.

In the present paper, we will study even-parity perturbations in the self-accelerating branch of MTBG.
In order to simplify the analysis of the even-parity perturbations, we will focus on the Schwarzschild background solutions written in the spatially-flat coordinates, where the effective cosmological constants are tuned to zero, instead of the Schwarzschild-de~Sitter solutions.
By construction, for $\ell \geq 2$ there should be two propagating DOFs. 
We also expect the appearance of a number of instantaneous modes which obey a set of elliptic differential equations on each constant time hypersurface.
To make the analysis of the $\ell \geq 1$ modes explicit,  we will set the ratio of the lapse functions between the two sectors to be unity.
In this case the effective graviton mass terms in the equations of motion of the perturbations vanish.
We will also assume that the two gravitational radii of the Schwarzschild metrics in both the sectors coincide.
We call this case the {\it effectively massless} case.
Within these assumptions, we will be able to reduce the set of the perturbation equations, and finally identify the two propagating modes and four instantaneous modes.

For the $\ell=0$ and $1$ modes, we will show that there is no propagating DOF as expected from the structure of MTBG, and will present the exact analytic solutions for the perturbations of the two spacetime metrics and the Lagrange multipliers.
For these modes, we will show that under the suitable boundary conditions all the free functions of time can be fixed.
For $\ell=0$, the solution for the perturbations can be absorbed by the redefinition of the two gravitational radii of the Schwarzschild solution.
We also find that all the components of the $\ell=1$ perturbations vanish by imposing the regularity of the physical and fiducial metrics at spatial infinity.

The structure of this paper is as follows:
In Sec.~\ref{sec2}, we briefly review MTBG.
In Sec.~\ref{sec3}, we introduce the even-parity perturbations about the Schwarzschild solution written in the spatially-flat coordinates in the self-accelerating branch of MTBG.
In Sec.~\ref{sec4}, we investigate the monopolar perturbations with $\ell=0$ and relate the solution to the nonlinear analysis of the time-dependent spherically symmetric solutions in the spatially-flat coordinates.
In Sec.~\ref{sec5}, we investigate the dipolar perturbations with $\ell=1$ within the effectively massless case and find the boundary conditions to fix all the free functions of time at the spatial infinities.
In Sec. \ref{sec6}, we investigate the higher multipolar perturbations with $\ell\geq 2$ in the effectively massless case and identify the two propagating and four instantaneous modes.
The last section \ref{sec7} is devoted to giving a brief summary and conclusion.

\section{Minimal theory of bigravity and Minkowski vacua}
\label{sec2}

\subsection{Theory}

We start with the Arnowitt-Deser-Misner (ADM) decomposition of the physical and fiducial metrics, $g_{\mu\nu}$ and $f_{\mu\nu}$,  respectively,
\begin{eqnarray}
\label{adm_decomposition}
g_{\mu\nu}dx^\mu dx^\nu &=&-N^2 dt^2+\gamma_{ij}\left(dx^i+N^i dt\right) \left(dx^j+N^j dt\right),
\nonumber
\\
f_{\mu\nu}dx^\mu dx^\nu &=&- M^2 dt^2+\phi_{ij}\left(dx^i+M^i dt\right)\left(dx^j+M^j dt\right),
\end{eqnarray}
where $x^\mu=(t,x^i)$ represents the coordinates of the four-dimensional spacetime with  $t$ and $x^i$ being the temporal coordinate and the coordinates of the three-dimensional spaces, respectively.
$N$, $N^i$, and $\gamma_{ij}$ represent the lapse function, shift vector, and three-dimensional spatial metric in the physical sector, and $M$, $M^i$, and $\phi_{ij}$ represent the corresponding quantities in the fiducial sector.

In the unitary gauge, the action of MTBG \cite{DeFelice:2020ecp,Minamitsuji:2022vfv,Minamitsuji:2023lvi} is then given by
\begin{eqnarray}
\label{action}
S
&=&
\frac{1}{2\kappa^2}
\int d^4x 
\left(
{\cal L}_g
\left[N,N^i,\gamma_{ij}; M,M^i,\phi_{ij};
\lambda, {\bar \lambda}, \lambda^i
\right]
+
{\cal L}_m
\left[
N,N^i,\gamma_{ij};
M,M^i,\phi_{ij};
\Psi
\right]
\right),
\end{eqnarray}
where
$\kappa^2=8\pi G$ represents the gravitational constant in the physical sector, ${\cal L}_g$ and ${\cal L}_m$ represent the gravitational and matter parts of the Lagrangian,  respectively,
$\lambda$, ${\bar \lambda}$, and $\lambda^i$ describe the two scalar and one spatial-vector Lagrange multipliers which are associated with the second-class constraints necessary to reduce the number of propagating DOFs to four in the original Hamiltonian formulation, and $\Psi$ represents the matter fields.
The gravitational Lagrangian ${\cal L}_g$ of MTBG is further decomposed into the `precursor' and `constraint' parts as 
\begin{eqnarray}
{\cal L}_g
&=&
{\cal L}_{\rm pre}
\left[
N,N^i,\gamma_{ij};
M,M^i,\phi_{ij}
\right]
+
{\cal L}_{\rm con}
\left[N,N^i,\gamma_{ij};
M,M^i,\phi_{ij};
\lambda, {\bar \lambda}, \lambda^i
\right],
\end{eqnarray}
with
\begin{eqnarray}
\label{lag2}
{\cal L}_{\rm pre}
&:=&
\sqrt{-g}R[g]
+
\alpha^2
\sqrt{-f} R[f]
-
m^2
\left(
N\sqrt{\gamma}{\cal H}_0
+
M\sqrt{\phi}\tilde{\cal H}_0
\right),
\\
\label{lag22}
{\cal L}_{\rm con}
&:=&
\sqrt{\gamma}
\alpha_{1\gamma}
\left(
\lambda
+
\Delta_\gamma
{\bar \lambda}
\right)
+
\sqrt{\phi}
\alpha_{1\phi}
\left(
\lambda
-
\Delta_\phi
{\bar \lambda}
\right)
+
\sqrt{\gamma}
\alpha_{2\gamma}
\left(
\lambda
+
\Delta_\gamma
 {\bar\lambda}
\right)^2
+
\sqrt{\phi}
\alpha_{2\phi}
\left(
\lambda
-\Delta_\phi {\bar\lambda}
\right)^2
\nonumber\\
&-&
m^2
\left[
\sqrt{\gamma}
  U^i{}_k D_i \lambda^k
-\beta 
\sqrt{\phi}
{\tilde U}_k{}^i {\tilde D}_i \lambda^k
\right],
\end{eqnarray}
and 
\begin{eqnarray}
\alpha_{1\gamma}
&:=&
-m^2
 U^p{}_q K^q{}_p,
\quad 
\alpha_{1\phi}
:=
m^2 {\tilde U}^p{}_q
    \Phi^q{}_p,
\nonumber\\
\alpha_{2\gamma}
&:=&
\frac{m^4 }{4N}
\left(
U^p{}_q-\frac{1}{2}U^k{}_k \delta^p{}_q
\right)
U^q{}_p,
\quad 
\alpha_{2\phi}
:=
\frac{m^4 }{4M\alpha^2}
\left(
{\tilde U}_q{}^p
-\frac{1}{2}{\tilde U}_k{}^k \delta_q{}^p
\right)
{\tilde U}_p{}^q,
\end{eqnarray}
where the constant $\alpha$ represents the ratio of the gravitational constants between the two sectors, $m$ is a parameter with mass dimension one which can be regarded as the graviton mass, $\beta$ is a constant, $\gamma:={\rm det} (\gamma_{ij})$ and $\phi:={\rm det} (\phi_{ij})$ are the determinants of the two three-dimensional spatial metrics $\gamma_{ij}$ and $\phi_{ij}$, respectively.
We also note that $K^q{}_p=\gamma^{qr}K_{rp}$ and $\Phi^q{}_p=\phi^{qr}\Phi_{rp}$, where $K_{ij}$ and $\Phi_{ij}$ represent the extrinsic curvature tensors on each constant time hypersurface in the physical and fiducial sectors, respectively.
Furthermore, ${\cal H}_0$
and $\tilde{\cal H}_0$ are defined by
$ {\cal H}_0 := \sum_{n=0}^3 c_{4-n} e_n( {\cal K})$ and
$ \tilde{\cal H}_0 := \sum_{n=0}^3 c_{n} e_n( \tilde{\cal K})$ with
\begin{eqnarray}
&&
e_0({\cal K})
=1,
\quad
e_1({\cal K})
=
\left[
{\cal K}
\right],
\quad 
e_{2} ({\cal K})
=
\frac{1}{2}
\left(
\left[
{\cal K}
\right]^2
-
\left[
{\cal K}^2
\right]
\right),
\quad
e_{3} ({\cal K})
=
{\rm det} 
({\cal K}),
\nonumber\\
&&
e_0(\tilde{\cal K})
=1,
\quad
e_1(\tilde{\cal K})
=
\left[
\tilde{\cal K}
\right],
\quad 
e_{2} (\tilde{\cal K})
=
\frac{1}{2}
\left(
\left[
\tilde{\cal K}
\right]^2
-
\left[
\tilde{\cal K}^2
\right]
\right),
\quad
e_{3} (\tilde{\cal K})
=
{\rm det} 
(\tilde{\cal K}),
\end{eqnarray}
with ${\cal K}^i{}_k$ and $\tilde{\cal K}_k{}^i$
characterized by ${\cal K}^i{}_k{\cal K}^k{}_j={\gamma}^{ik}\phi_{kj}$ and ${\cal {\tilde K}}_j{}^k{\cal {\tilde K}}_k{}^i=\gamma_{jk}{\phi}^{ki}$.
$\Delta_\gamma :=\gamma^{ij} D_i D_j$ and $\Delta_\phi :=\phi^{ij} {\tilde D}_i {\tilde D}_j$ represent the Laplacian operators in the physical and fiducial sectors, respectively, where $D_i$ and ${\tilde D}_i$ are the covariant derivatives associated with the spatial metrics $\gamma_{ij}$ and $\phi_{ij}$.
The spatial tensors $U^i{}_j$ and ${\tilde U}{}_j{}^i$ are, respectively, defined by 
\begin{eqnarray}
U^i{}_j
:=
\frac{1}{2}
\sum_{n=1}^3 c_{4-n}
\left(
U_{(n)}{}^i{}_j
+\gamma^{ik}\gamma_{j\ell}
U_{(n)}{}^\ell{}_k
\right),
\quad
{\tilde U}_j{}^i
:=
\frac{1}{2}
\sum_{n=1}^3 c_{n}
\left(
{\tilde U}_{(n)j}{}^i
+\phi^{ik}\phi_{j\ell}
{\tilde U}_{(n)k}{}^\ell
\right),
\end{eqnarray}
with
$U_{(n)}{}^i{}_k:=\frac{\partial e_n ({\cal K})} {\partial {\cal K}^k{}_i}$, ${\tilde U}_{(n)k}{}^i :=\frac{\partial e_n ({\cal {\tilde K}})}{\partial \tilde{\cal K}^k{}_i}$, and $c_j$ ($j=0,1,2,3,4$) being dimensionless coupling constants.
In this paper, we will focus on the vacuum case and set ${\cal L}_m=0$ in Eq.~\eqref{action}.

\subsection{Minkowski vacua in the self-accelerating branch}

Before considering the BH solutions, we briefly review de~Sitter and Minkowski solutions in the spatially-flat, homogeneous and isotropic FLRW metrics,
which are respectively given by 
\begin{eqnarray}
\label{generalfg_ds}
g_{\mu\nu}
dx^\mu dx^\nu
&=&
-dt^2
+
\left(dr-r\frac{\dot{a}(t)}{a(t)} dt\right)^2
+r^2 
\left(
 d\theta^2
+\sin^2\theta d\varphi^2
\right),
\nonumber\\
f_{\mu\nu}
dx^\mu dx^\nu
&=&
C_0^2
\left[
-b^2C_m(t)^2 dt^2
+\left(dr-r\frac{\dot{a}_f(t)}{a_f(t)} C_m(t) dt\right)^2
+ r^2 
\left(
 d\theta^2
+\sin^2\theta d\varphi^2
\right)
\right],
\end{eqnarray}
where $a(t)$ and $a_f(t)$ represent the scale factors in the physical and fiducial sectors, respectively, $C_m(t)$ is a function of the time $t$, and $b$ is a constant that characterizes the relative light cone aperture in the fiducial sector.
In the presence of a nontrivial form of $C_m(t)$, the time coordinate $t$ cannot be the proper time in the fiducial sector.
The general ansatz for the Lagrange multipliers is given by 
\begin{eqnarray}
\label{spherical_lagrange}
\lambda=\lambda(t,r),
\qquad 
{\bar \lambda}={\bar \lambda}(t,r),
\qquad 
\lambda^r=\lambda^r(t,r),
\qquad 
\lambda^\theta=\lambda^\varphi=0.
\end{eqnarray}
In order for the equations of motion for $\lambda$, ${\bar \lambda}$, and $\lambda^r$ to be automatically satisfied,
we impose the condition for the self-accelerating branch \cite{DeFelice:2020ecp,Minamitsuji:2022vfv,Minamitsuji:2023lvi} given by 
\begin{eqnarray}
c_3+2C_0 c_2+C_0^2c_1=0.
\label{sa}
\end{eqnarray}
From the equations of motion for the lapse functions $N$ and $M$, i.e., the Friedmann equations in both the sectors,  we obtain the solutions for the scale factors in the physical and fiducial sectors given by 
\begin{eqnarray}
\label{de_Sitter_solutions}
&&
a(t)
=
a_0\,
{\rm exp}
\left[
     \sqrt{\frac{\Lambda_g}{3}}
t
\right],
\qquad
a_f(t)=
a_{f,0}\,
{\rm exp}
\left[
     b C_0
     \sqrt{\frac{\Lambda_f}{3}}
t
\right],
\end{eqnarray}
where $a_0$ and $a_{0,f}$ are integration constants,
and  the effective cosmological constants in both the sectors are related to the parameters in the Lagrangian of MTBG \eqref{lag2} by 
\begin{eqnarray}
\label{Lambdas}
&&
\Lambda_{g}=\frac{m^{2} \left(c_4-2 C_{0}^{3} c_{1}-3 C_{0}^{2} c_{2}\right)}{2},
\qquad
\Lambda_{f}=\frac{\left(C_{0}^{2} c_{0}+2 C_{0} c_{1}+c_{2}\right) m^{2}}{2 C_{0}^{2} {\alpha}^{2}}.
\end{eqnarray}
The remaining metric equations of motion provide the general solutions for the Lagrange multipliers
\begin{eqnarray}
\label{lambda_sol_cosmology}
\lambda
&=&
C_\lambda(t)
+
\frac{1}{C_0^2}
\frac{\sqrt{c_0C_0^2+2C_0c_1+c_2}
+
C_0^2
\sqrt{-2C_0^3c_1-3C_0^2c_2+c_4}\alpha}
        {\sqrt{c_0C_0^2+2C_0c_1+c_2}
-
\sqrt{-2C_0^3c_1-3C_0^2c_2+c_4}\alpha}
\left(
{\bar \lambda}''
+
\frac{2}{r}
{\bar \lambda}'
\right),
\\
\lambda^r
&=&
-
\frac{\sqrt{c_0C_0^2+2C_0c_1+c_2}
-
\sqrt{-2C_0^3c_1-3C_0^2c_2+c_4}\alpha }
       {\sqrt{6}\alpha (-1+C_0\beta)}
mr C_{\lambda}(t),
\end{eqnarray}
while ${\bar \lambda}$ remains undetermined.
Imposing the regularity of $\lambda^r$ as $r\to \infty$ yields
\begin{eqnarray}
C_\lambda(t)=0,
\end{eqnarray}
which leads to $\lambda^r=0$ and, from Eq. \eqref{lambda_sol_cosmology},
\begin{eqnarray}
\lambda
&=&
\frac{1}{C_0^2}
\frac{\sqrt{c_0C_0^2+2C_0c_1+c_2}
+
C_0^2
\sqrt{-2C_0^3c_1-3C_0^2c_2+c_4}\alpha}
        {\sqrt{c_0C_0^2+2C_0c_1+c_2}
-
\sqrt{-2C_0^3c_1-3C_0^2c_2+c_4}\alpha}
\left(
{\bar \lambda}''
+
\frac{2}{r}
{\bar \lambda}'
\right).
\end{eqnarray}
Thus, ${\lambda}$ and $\bar{\lambda}$ are not determined by the background equations of motion.
\footnote{At higher-order the metric may depend on ${\bar\lambda}$ and/or $\bar{\bar\lambda}$. If this is the case then ${\bar\lambda}$ and/or $\bar{\bar\lambda}$ may be fixed by suitable boundary conditions for higher-order perturbations. However, this is beyond the scope of the present paper.}

In the limit of the vanishing effective cosmological constants $\Lambda_g=0$ and $\Lambda_f=0$, which from Eq.~\eqref{Lambdas} are explicitly given by 
\begin{eqnarray}
c_4-2 C_{0}^{3} c_{1}-3 C_{0}^{2} c_{2}=0,
\qquad
C_{0}^{2} c_{0}+2 C_{0} c_{1}+c_{2}=0,
\label{flat}
\end{eqnarray}
the paired de~Sitter solutions \eqref{de_Sitter_solutions} can smoothly reduce to the paired Minkowski solutions
\begin{eqnarray}
\label{generalfg_ds2}
g_{\mu\nu}
dx^\mu dx^\nu
&=&
-dt^2
+
dr^2
+r^2 
\left(
 d\theta^2
+\sin^2\theta d\varphi^2
\right),
\nonumber\\
f_{\mu\nu}
dx^\mu dx^\nu
&=&
C_0^2
\left[
-b^2C_m(t)^2 dt^2
+dr^2
+ r^2 
\left(
 d\theta^2
+\sin^2\theta d\varphi^2
\right)
\right].
\end{eqnarray}
We note that the nontrivial form of $C_m(t)$ represents a Minkowski vacuum in the fiducial sector which is different from that in the physical sector.
In order that the physical and fiducial sectors share the same Minkowski vacua, we have to set $C_m(t)=1$.
In the next sections,  we assume that in the two asymptotic regions of the paired Schwarzschild solution written in the spatially-flat coordinates, the physical and fiducial sectors share the same Minkowski vacua.

\section{Even-parity perturbations of Schwarzschild solutions}
\label{sec3}

Under the ADM decomposition \eqref{adm_decomposition}, the perturbed static and spherically symmetric spacetimes written in the spatially-flat coordinates can be described as
\begin{eqnarray}
g_{\mu\nu}dx^\mu dx^\nu
&=&
-N^2 dt^2
+\gamma_{rr}
\left(dr+N^rdt\right)^2
+
2\gamma_{ra}
\left(dr+N^rdt\right)
\left(d\theta^a+N^a dt\right)
+
\gamma_{ab}
\left(d\theta^a+N^a dt\right)
\left(d\theta^b+N^b dt\right),
\nonumber
\\
f_{\mu\nu}dx^\mu dx^\nu
&=&
-b^2 M^2 dt^2
+\phi_{rr}
\left(dr+M^rdt\right)^2
+
2\phi_{ra}
\left(dr+M^rdt\right)
\left(d\theta^a+M^a dt\right)
\nonumber\\
&&
+
\phi_{ab}
\left(d\theta^a+M^a dt\right)
\left(d\theta^b+M^b dt\right),
\end{eqnarray}
where $r$ is the radial coordinate, $\theta^a=(\theta,\varphi)$ represents the coordinates along the unit two-sphere, 
and 
\begin{eqnarray}
N&=&N_{(0)}(r)\left(1+ \sum_{\ell,m} n_0(t,r) Y_{\ell m}(\theta^a)\right),
\qquad
N^r =N^r_{(0)}(r)\left(1+\sum_{\ell,m} n_1 (t,r)Y_{\ell m}(\theta^a)\right),
\nonumber\\
\gamma_{rr}
&=&1+\sum_{\ell,m} n_2(t,r) Y_{\ell m}(\theta^a),
\quad
N^a
=
\sum_{\ell,m}
h_t(t,r)
 \theta^{ab}{\tilde \nabla}_b Y_{\ell m}(\theta^a),
\qquad
\gamma_{ra}
=
r
\sum_{\ell,m}
h_r (t,r)
{\tilde \nabla}_a Y_{\ell m}(\theta^a),
\nonumber
\\
\gamma_{ab}
&=&
r^2
\left[
\theta_{ab}
+
\sum_{\ell,m}
h_1 (t,r)
\theta_{ab}Y_{\ell m}(\theta^a)
+
\sum_{\ell,m}
h_2(t,r)
{\tilde \nabla}_a {\tilde \nabla}_b Y_{\ell m}(\theta^a)
\right],
\nonumber\\
M&=&
C_0
M_{(0)}(r)
\left(
1+\sum_{\ell,m} m_0(t,r) Y_{\ell m} (\theta^a)
\right),
\qquad
M^r =M^r_{(0)}(r)\left(1+\sum_{\ell,m} m_1(t,r) Y_{\ell m}(\theta^a)\right),
\nonumber\\
\phi_{rr}
&=&
C_0^2
\left(
1+\sum_{\ell,m}m_2(t,r) Y_{\ell m}(\theta^a)\right),
\quad
M^a
=
\sum_{\ell,m}
k_t (t,r)
\theta^{ab}
{\tilde \nabla}_b Y_{\ell m}(\theta^a),
\qquad
\phi_{ra}
=
C_0^2
r
\sum_{\ell,m}
k_r (t,r)
{\tilde \nabla}_a Y_{\ell m}(\theta^a),
\nonumber\\
\phi_{ab}
&=&
C_0^2
r^2
\left[
\theta_{ab}
+
\sum_{\ell,m}
k_1 (t,r)\theta_{ab}Y_{\ell m}(\theta^a)
+
\sum_{\ell,m}
k_2(t,r)
{\tilde \nabla}_a {\tilde \nabla}_b 
Y_{\ell m}(\theta^a)
\right].
\end{eqnarray}
Here, $C_0>0$ is a constant fixed by solving the background equations of motion, $Y_{\ell m}(\theta^a)$ represents spherical harmonics with the multipole and magnetic moments $(\ell,m)$ with $-\ell\leq m\leq \ell$, and ${\tilde \nabla}_a$ represents the covariant derivative with respect to the metric of the unit two-sphere $\theta_{ab}$, respectively.
Because of the degeneracy between the different $m$ modes for the same $\ell$, without loss of generality, we may choose $m=0$  so that $Y_{\ell m}=P_\ell (\cos\theta)$.
The parameter $b\,(>0)$ measures the difference in the time passing in the physical and fiducial sectors, which acquires a nontrivial physical significance in MTBG where the two copies of the four-dimensional diffeomorphism invariance is broken down to the joint three-dimensional one.
$N_{(0)} (r)$ and $N^r_{(0)}(r)$ represent the background lapse function and shift vector in the physical sector, while $M_{(0)} (r)$ and $M^r_{(0)}(r)$ represent the corresponding quantities in the fiducial sector, respectively.
For each of the $\ell$ modes, $n_0$, $n_1$, $n_2$, $h_t$, $h_r$, $h_1$, and $h_2$ represent the $(t,r)$ part of the metric perturbations in the physical sector, and $m_0$, $m_1$, $m_2$, $k_t$, $k_r$, $k_1$, and $k_2$ represent the $(t,r)$ part of the metric perturbations in the fiducial sector.

In the self-accelerating branch of MTBG satisfying Eq.~\eqref{sa}, the Schwarzschild-de~Sitter solution written in the spatially-flat coordinates is given by 
\begin{eqnarray}
\label{metric_g}
g_{\mu\nu}dx^\mu dx^\nu
&=&
-
N_{(0)}(r)^2
dt^2
+
\left(dr+N_{(0)}^r(r) dt\right)^2
+r^2\theta_{ab}d\theta^a d\theta^b,
\nonumber\\
\label{metric_f}
f_{\mu\nu}dx^\mu dx^\nu
&=&
C_0^2
\Big[
-
b^2
M_{(0)}(r)^2
dt^2
+
\left(dr+M_{(0)}^r(r) dt\right)^2
+r^2\theta_{ab}d\theta^a d\theta^b
\Big],
\end{eqnarray}
with
\begin{eqnarray}
\label{outside2}
&&
N_{(0)}(r)
=
M_{(0)}(r)
=1,
\qquad 
N_{(0)}^{r}(r)=\sqrt{ \frac{\Lambda_{g} r^{2}}{3}+\frac{r_g}{r}},
\qquad 
M_{(0)}^r(r)=\sqrt{\frac{r^{2} C_{0}^{2} b^{2} \Lambda_{f}}{3}+\frac{r_{f}}{r}},
\end{eqnarray}
where 
$\Lambda_{g}$ and $\Lambda_{f}$
are the effective cosmological constants given by Eq.~\eqref{Lambdas}~\cite{Minamitsuji:2022vfv,Minamitsuji:2023lvi}.

In the rest, we focus on the Schwarzschild solution obtained in the limit of the vanishing effective cosmological constants $\Lambda_{g}=0$ and $\Lambda_{f}=0$, under the conditions explicitly given by Eq.~\eqref{flat}.
The background metric solution \eqref{outside2} in this limit reduces to 
\begin{eqnarray}
\label{Schwarzschilld_metrics}
N_{(0)}(r)
=
M_{(0)}(r)
=1,
\qquad 
N_{(0)}^{r}(r)=\sqrt{\frac{r_g}{r}},
\qquad 
M_{(0)}^r(r)=\sqrt{\frac{r_{f}}{r}},
\end{eqnarray}
where $r_g$ and $r_f$ then correspond to the gravitational radii of the Schwarzschild spacetimes.
We will assume that $C_0c_1+c_2\neq 0$.
As mentioned previously, we also call the case of $b=1$ (as well as $r_f=r_g$) the {\it effectively massless case}, where the effective graviton mass terms in the equations of motion of perturbations vanish in both the odd- and even-parity sectors \cite{Minamitsuji:2023lvi}.

Because of the spherical symmetry, the background part of the angular components of the vector Lagrange multiplier trivially vanish, $\lambda^\theta=\lambda^\varphi=0$.
We also choose the trivial solution for the remaining components of the background Lagrange multipliers
\begin{eqnarray}
\label{tri_multi}
\lambda=0,
\qquad 
{\bar \lambda}=0,
\qquad 
{\lambda}^r=0,
\end{eqnarray}
which is compatible with the background equations of motion.
In general, the background solution for ${\bar \lambda}$ may be a solution for the Laplace equation in three-dimensional flat space.
However, since in the Lagrangian \eqref{lag2} the ${\bar \lambda}$ dependence appears through the spatial Laplacian operators $\Delta_\gamma {\bar\lambda}$ and $\Delta_\phi {\bar\lambda}$,
i.e., in our background case the Laplacian operator in the three-dimensional flat space acting on ${\bar \lambda}$,
 a solution of the Laplace equation in the three-dimensional flat space does not contribute to the background dynamics. 
Thus, without loss of generality, we may set ${\bar\lambda}=0$.
On top of the trivial solution~\eqref{tri_multi}, we consider the even-parity perturbations of the Lagrange multipliers given by 
\begin{eqnarray}
\label{pert_lag}
&&
\lambda=\sum_{\ell,m}\lambda_0 (t,r)Y_{\ell m} (\theta^a),
\qquad
{\bar\lambda}=\sum_{\ell,m}\lambda_1(t,r) Y_{\ell m}(\theta^a),
\nonumber
\\
&&
\lambda^r= \sum_{\ell,m}\lambda_2 (t,r)Y_{\ell m}(\theta^a),
\qquad
\lambda^a=\sum_{\ell,m}\lambda_3(t,r) \theta^{ab}{\tilde \nabla}_b Y_{\ell m}(\theta^a).
\end{eqnarray}

\subsection{The $\ell\geq 2$ modes}

For the $\ell \geq 2 $ modes, the perturbed physical and fiducial metrics in the even-parity sectors are, respectively, given by 
\begin{eqnarray}
\label{metric_g_ell}
g_{\mu\nu}dx^\mu dx^\nu
&=&
-
N_{(0)}(r)^2
dt^2
+
\left(dr+N_{(0)}^r(r) dt\right)^2
+r^2\theta_{ab}d\theta^a d\theta^b
\nonumber\\
&+&
\sum_{\ell\geq 2,m}
\Big\{
\left[
-2n_0(t,r) N_{(0)}(r)^2
+N^r_{(0)}(r)^2
 \left(n_2(t,r)+2n_1(t,r)\right)
\right]
Y_{\ell m}(\theta^a) dt^2
\nonumber\\
&+&
2N^r_{(0)}(r)
\left(
n_1(t,r)
+
n_2(t,r)
\right)
Y_{\ell m}(\theta^a)
dt dr
+
n_2(t,r) Y_{\ell m} (\theta^a)
dr^2
\nonumber\\
&+&
2r \left(r h_t(t,r)+N^r_{(0)}(r)h_r(t,r)\right) {\tilde \nabla}_a Y_{\ell m} dt d\theta^a
+
2r  h_r (t,r) {\tilde \nabla}_a Y_{\ell m} (\theta^a) dr d\theta^a
\nonumber\\
&+&
r^2
\left(
h_1(t,r) \theta_{ab} Y_{\ell m} (\theta^a)
+
h_2(t,r)
{\tilde \nabla}_a {\tilde \nabla}_b Y_{\ell m} (\theta^a)
\right)
d\theta^a
d\theta^b
\Big\},
\\
\label{metric_f_ell}
f_{\mu\nu}dx^\mu dx^\nu
&=&
C_0^2
\Big[
-
b^2
M_{(0)}(r)^2
dt^2
+
\left(dr+M_{(0)}^r(r) dt\right)^2
+r^2\theta_{ab}d\theta^a d\theta^b
\nonumber\\
&+&
\sum_{\ell\geq 2,m}
\Big\{
\left(-2m_0 b^2 M_{(0)}(r)^2
+M^r_{(0)}(r)^2
 \left((m_2(t,r)+2m_1(t,r)\right)
\right)
Y_{\ell m}(\theta^a)dt^2
\nonumber\\
&+&
2M^r_{(0)}(r)
\left(m_1(t,r)+m_2(t,r)\right)
Y_{\ell m} (\theta^a)
dt dr
+
m_2(t,r) Y_{\ell m} (\theta^a)
dr^2
\nonumber\\
&+&
2r \left(r k_t(t,r)+M^r_{(0)}(r)k_r(t,r)\right) {\tilde \nabla}_a Y_{\ell m} (\theta^a) dt d\theta^a
+
2 rk_r(t,r)  {\tilde \nabla}_a Y_{\ell m}  (\theta^a) dr d\theta^a
\nonumber\\
&+&
r^2
\left(
k_1(t,r) \theta_{ab} Y_{\ell m} (\theta^a)
+
k_2(t,r)
{\tilde \nabla}_a {\tilde \nabla}_b Y_{\ell m}(\theta^a)
\right)
d\theta^a
d\theta^b
\Big\}
\Big].
\end{eqnarray}
The perturbed three-dimensional metrics in the even-parity sector are, respectively, given by
\begin{eqnarray}
\gamma_{ij}dy^i dy^j
&=&
dr^2
+
r^2\theta_{ab}
d\theta^a 
d\theta^b
\nonumber \\
&+&
\sum_{\ell\geq2,m}
\Big[
n_2(t,r) Y_{\ell m} (\theta^a)dr^2
+
2r h_r(t,r) {\tilde \nabla}_a Y_{\ell m}(\theta^a) dr d\theta^a
\nonumber\\
&+&
r^2
\left(
h_1(t,r)\theta_{ab} Y_{\ell m}(\theta^a)
+
h_2(t,r)
{\tilde \nabla}_a {\tilde \nabla}_b Y_{\ell m}(\theta^a)
\right)
d\theta^a
d\theta^b
\Big],
\nonumber
\\
\phi_{ij}dy^i dy^j
&=&
C_0^2
\Big\{
dr^2
+
r^2\theta_{ab}d\theta^a d\theta^b
\nonumber\\
&+&
\sum_{\ell\geq 2,m}
\Big[
m_2 (t,r)Y_{\ell m} (\theta^a)dr^2
+
2r k_r (t,r){\tilde \nabla}_a Y_{\ell m}(\theta^a) dr d\theta^a
\nonumber\\
&+&
r^2
\left(
k_1(t,r)
\theta_{ab} Y_{\ell m}(\theta^a)
+
k_2(t,r)
{\tilde \nabla}_a {\tilde \nabla}_b Y_{\ell m}(\theta^a)
\right)
d\theta^a
d\theta^b
\Big]
\Big\}.
\end{eqnarray}
Under the spatial gauge-transformation $t\to t$ and $x^i \to x^i +\xi^i(t,x^i) $ with
\begin{eqnarray}
\label{gauge_transformation}
\xi^r= 
\sum_{\ell\geq 2,m}
\Xi_r (t,r)Y_{\ell m}(\theta^a),
\qquad 
\xi^a
=
\sum_{\ell\geq 2,m}
\Xi_1(t,r)
\theta^{ab}
{\tilde \nabla}_b Y_{\ell m}(\theta^a),
\end{eqnarray}
the metric perturbations transform as 
\begin{eqnarray}
\label{rule_g_ell1}
{\bar \delta} n_0(t,r)
&=&
-\frac{N_{(0)} '(r)}
         {N_{(0)} (r)}
\Xi_r(t,r),
\qquad
{\bar \delta} n_1(t,r)
=
-\frac{N^r_{(0)} {}'(r)}
         {N^r_{(0)} (r)}
\Xi_r(t,r)
+
\Xi_r{}'(t,r)
-
\frac{1}
       {N^r_{(0)}(r)}
\dot{\Xi}_{r}(t,r),
\nonumber
\\
{\bar\delta} n_2(t,r)
&=&
-2
\Xi_r'(t,r),
\qquad 
{\bar\delta} 
h_t(t,r)
=
N^r_{(0)}(r)
\Xi_1'(t,r)
-
\dot{\Xi}_1(t,r),
\qquad 
{\bar\delta} 
h_r(t,r)
=
-
\frac{
\Xi_r(t,r)
+r^2 \Xi_1'(t,r)}
{r},
\nonumber\\
{\bar\delta} 
h_1(t,r)
&=&
-\frac{2\Xi_r(t,r)}{r},
\qquad
{\bar\delta} 
h_2(t,r)
=
-2\Xi_1(t,r),
\end{eqnarray}
and 
\begin{eqnarray}
\label{rule_f_ell1}
{\bar \delta} m_0
(t,r)
&=&
-\frac{M_{(0)}'(r)}
         {M_{(0)} (r)}
\Xi_r(t,r),
\qquad
{\bar \delta} m_1
(t,r)
=
-\frac{M^r_{(0)}{}'(r)}
         {M^r_{(0)} (r)}
\Xi_r(t,r)
+
\Xi_r'(t,r)
-
\frac{1}{M^r_{(0)}(r)}
\dot{\Xi}_{r}(t,r),
\nonumber
\\
\qquad
{\bar\delta} 
m_2
(t,r)
&=&
-2\Xi_r'(t,r),
\qquad 
{\bar\delta} 
k_t
(t,r)
=
M^r_{(0)}(r)
\Xi_1'(t,r)
-
\dot{\Xi}_1(t,r),
\qquad 
{\bar\delta} 
k_r
(t,r)
=
-
\frac{\Xi_r(t,r)+ r^2 \Xi_1'(t,r)}{r},
\nonumber\\
{\bar\delta} 
k_1
(t,r)
&=&
-\frac{2\Xi_r(t,r)}{r},
\qquad
{\bar\delta} 
k_2
(t,r)
=
-2\Xi_1(t,r),
\end{eqnarray}
where $\bar\delta$ represents the difference between the perturbed quantities before and after the gauge transformation.
For each of the $\ell\geq 2$ modes, among fourteen metric variables for the even parity perturbations, two of them can be eliminated by choosing $\Xi_r$ and $\Xi_1$.
Later, we move to the gauge 
\begin{eqnarray}
h_1(t,r)=h_2(t,r)=0,
\label{h1h20}
\end{eqnarray}
which will fix the two gauge functions $\Xi_r$ and $\Xi_1$ completely.

\subsection{The $\ell=0$ mode}

For the $\ell=0$ mode,
using $Y_{\ell m}=P_0(\cos\theta)=1$,
the metric perturbations can be written as 
\begin{eqnarray}
\label{metric_g_ell0}
g_{\mu\nu}dx^\mu dx^\nu
&=&
-
N_{(0)}(r)^2
dt^2
+
\left(dr+N_{(0)}^r(r) dt\right)^2
+r^2\theta_{ab}d\theta^a d\theta^b
\nonumber\\
&+&
\left[
-2n_0(t,r) N_{(0)}(r)^2
+N^r_{(0)}(r)^2
 \left(n_2(t,r)+2n_1(t,r)\right)
\right]
dt^2
+2N^r_{(0)}(r)
\left(
n_1(r,r)+n_2(t,r)
\right)
dt dr
\nonumber\\
&+&
n_2 (t,r)
dr^2
+
r^2
h_1 (t,r)\theta_{ab} 
d\theta^a
d\theta^b,
\\
\label{metric_f_ell0}
f_{\mu\nu}dx^\mu dx^\nu
&=&
C_0^2
\Big\{
-
b^2
M_{(0)}(r)^2
dt^2
+
\left(dr+M_{(0)}^r(r) dt\right)^2
+r^2\theta_{ab}d\theta^a d\theta^b
\nonumber\\
&+&
\left[-2m_0(t,r) b^2 M_{(0)}(r)^2
+M^r_{(0)}(r)^2
 \left(m_2(t,r)+2m_1(t,r)\right)\right]
dt^2
+2M^r_{(0)}(r)
\left(m_1(t,r)+m_2(t,r)\right)
dt dr
\nonumber\\
&+&
m_2 (t,r)
dr^2
+
r^2
k_1 (t,r)
\theta_{ab} 
d\theta^a
d\theta^b
\Big\}.
\end{eqnarray}
Under the spatial gauge transformation for $\ell=0$, $t\to t$ and $x^i \to x^i +\xi^i(t,x^i) $ with $\xi^r= \Xi_r (t,r)$ and $\xi^a=0$, the metric perturbations transform as 
\begin{eqnarray}
{\bar \delta} n_0(t,r)
&=&
-\frac{N_{(0)}{}'(r)}
         {N_{(0)}(r)}
\Xi_r(t,r),
\qquad
{\bar \delta} n_1(t,r)
=
-\frac{N^r_{(0)}{}'(r)}
         {N^r_{(0)}(r)}
\Xi_r(t,r)
+
\Xi_r'(t,r)
-
\frac{1}{N^r_{(0)}(r)}
\dot{\Xi}_{r}(t,r),
\nonumber\\
{\bar\delta} n_2(t,r)
&=&
-2
\Xi_r'(t,r),
\qquad
{\bar\delta} 
h_1(t,r)
=
-\frac{2\Xi_r(t,r)}{r},
\end{eqnarray}
and 
\begin{eqnarray}
{\bar \delta} m_0
(t,r)
&=&
-\frac{(M_{(0)})'}
         {M_{(0)}}
\Xi_r(t,r),
\qquad
{\bar \delta} m_1
(t,r)
=
-\frac{M^r_{(0)}{}'(r)}
         {M^r_{(0)}(r)}
\Xi_r(t,r)
+
\Xi_r'(t,r)
-
\frac{1}{M^r_{(0)}(r)}
\dot{\Xi}_{r}(t,r),
\nonumber\\
{\bar\delta} 
m_2
(t,r)
&=&
-2\Xi_r'(t,r),
\quad 
{\bar\delta} 
k_1
(t,r)
=
-\frac{2\Xi_r(t,r)}{r}.
\end{eqnarray}
We will move to the gauge
\begin{eqnarray}
\label{gauge_ell0}
h_1(t,r)=0,
\end{eqnarray}
which completely fixes $\Xi_r$.
The even-parity perturbation of the Lagrange multipliers for $\ell =0$ is given by 
\begin{eqnarray}
\lambda=\lambda_0 (t,r),
\qquad
{\bar\lambda}=\lambda_1(t,r),
\qquad 
\lambda^r=\lambda_2 (t,r),
\qquad
\lambda^a=0.
\end{eqnarray}

\subsection{The $\ell=1$ mode}

For the $\ell=1$ mode, using $Y_{\ell m}=P_1(\cos\theta)=\cos\theta$, the two perturbed metrics can be expanded as 
\begin{eqnarray}
\label{metric_g}
g_{\mu\nu}dx^\mu dx^\nu
&=&
-
N_{(0)}(r)^2
dt^2
+
\left(dr+N_{(0)}^r(r) dt\right)^2
+r^2\theta_{ab}d\theta^a d\theta^b
\nonumber\\
&+&
\left[
-2n_0(t,r) N_{(0)}(r)^2
+N^r_{(0)}(r)^2
 \left(n_2(t,r)+2n_1(t,r)\right)
\right]
\cos\theta
dt^2
+2N^r_{(0)}(r)
\left(
n_1(t,r)+n_2(t,r)
\right)
\cos\theta
dt dr
\nonumber\\
&+&
n_2 (t,r)
\cos\theta
dr^2
-
2r \left(r h_t(t,r)+N^r_{(0)} (r)h_r(t,r)\right)  \sin\theta dt d\theta
\nonumber\\
&-&
2r  h_r(t,r) \sin\theta dr d\theta
+
r^2
\left(h_1(t,r)-h_2(t,r)\right)
\cos\theta
\cdot
\theta_{ab}
d\theta^a
d\theta^b,
\\
\label{metric_f}
f_{\mu\nu}dx^\mu dx^\nu
&=&
C_0^2
\Big\{
-
b^2
M_{(0)}(r)^2
dt^2
+
\left(dr+M_{(0)}^r(r) dt\right)^2
+r^2\theta_{ab}d\theta^a d\theta^b
\nonumber\\
&+&
\left[-2m_0(t,r) b^2 M_{(0)}(r)^2
+M^r_{(0)}(r)^2
 (m_2+2m_1)\right]
\cos\theta
dt^2
+2M^r_{(0)}(r)
\left(m_1(t,r)+m_2(t,r)\right)
\cos\theta
dt dr
\nonumber\\
&+&
m_2 (t,r)
\cos\theta
dr^2
-
2r \left(r k_t(t,r)+M^r_{(0)} (r)k_r(t,r)\right) \sin\theta dt d\theta
\nonumber\\
&-&
2 rk_r (t,r) \sin\theta dr d\theta
+
r^2
\left(
k_1(t,r)
-
k_2(t,r)
\right)
\cos\theta
\cdot
\theta_{ab}
d\theta^a
d\theta^b
\Big\}.
\end{eqnarray}
Under the spatial gauge-transformation $t\to t$ and $x^i \to x^i +\xi^i(t,x^i) $ for $\ell=1$ with  $\xi^r= \Xi_r (t,r)\cos\theta$ and $\xi^\theta=-\Xi_1(t,r)\sin\theta$, the metric perturbations transform as 
\begin{eqnarray}
{\bar \delta} n_0(t,r)
&=&
-\frac{N_{(0)}{}'(r)}
         {N_{(0)} (r)}
\Xi_r(t,r),
\qquad 
{\bar \delta} n_1(t,r)
=
-\frac{N^r_{(0)}{}'(r)}
         {N^r_{(0)} (r) }
\Xi_r(t,r)
+
\Xi_r'(t,r)
-
\frac{1}{N^r_{(0)}(r)}
\dot{\Xi}_{r}(t,r),
\nonumber
\\
{\bar\delta} n_2(t,r)
&=&
-2
\Xi_r'(t,r),
\qquad 
{\bar\delta} 
h_t(t,r)
=
N^r_{(0)}
\Xi_1'
-
\dot{\Xi}_1,
\nonumber\\
{\bar\delta} 
h_r(t,r)
&=&
-
\frac{
\Xi_r(t,r)
+r^2 \Xi_1'(t,r)}
{r},
\qquad 
{\bar\delta} 
(h_1(t,r)-h_2(t,r))
=
2\Xi_1(t,r)
-\frac{2\Xi_r(t,r)}{r},
\end{eqnarray}
and
\begin{eqnarray}
{\bar \delta} m_0(t,r)
&=&
-\frac{M_{(0)}{}'(r)}
         {M_{(0)}}
\Xi_r(t,r),
\qquad
{\bar \delta} m_1
(t,r)
=
-\frac{M^r_{(0)}{}'(r)}
         {M^r_{(0)} (r)}
\Xi_r(t,r)
+
\Xi_r'(t,r)
-
\frac{1}{M^r_{(0)}(r)}
\dot{\Xi}_{r}(t,r),
\nonumber\\
{\bar\delta} 
m_2
(t,r)
&=&
-2\Xi_r'(t,r),
\qquad 
{\bar\delta} 
k_t
(t,r)
=
M^r_{(0)}(r)
\Xi_1'(t,r)
-
\dot{\Xi}_1(t,r),
\nonumber
\\
{\bar\delta} 
k_r
(t,r)
&=&
-
\frac{\Xi_r(t,r)+ r^2 \Xi_1'(t,r)}{r},
\qquad
{\bar\delta} 
(k_1(t,r)-k_2(t,r))
=
2\Xi_1(t,r)
-\frac{2\Xi_r (t,r)}{r}.
\end{eqnarray}
Since $h_2(t,r)$ and $k_2(t,r)$ always appear as the combinations of $h_1(t,r)-h_2(t,r)$ and $k_1(t,r)-k_2(t,r)$ in Eqs.~\eqref{metric_g} and \eqref{metric_f}, respectively, we may set
\begin{eqnarray}
\label{h2k2}
h_2(t,r)=0,
\qquad 
k_2(t,r)=0.
\end{eqnarray}
By fixing $\Xi_r$ and $\Xi_1$ appropriately, for instance, we may choose the gauge
\begin{eqnarray}
\label{gauge_l1}
h_1(t,r)=0,
\qquad
h_r(t,r)=0.
\end{eqnarray}
The even-parity perturbation of the Lagrange multipliers for $\ell =1$
is given by 
\begin{eqnarray}
\lambda=\lambda_0 (t,r)\cos\theta,
\qquad
{\bar\lambda}=\lambda_1(t,r) \cos\theta,
\qquad 
\lambda^r=\lambda_2 (t,r)\cos\theta,
\qquad
\lambda^\theta=-\lambda_3(t,r) \sin\theta,
\qquad 
\lambda^\varphi=0.
\end{eqnarray}

\section{The solutions for the monopolar perturbations}
\label{sec4}

In this section, we focus on the $\ell =0$ mode.

\subsection{The case of the two copies of GR}

First, we focus on the case of the two copies of GR. This will help illustrate the case of MTBG, whose Lagrangian can be understood as a particular coupling between the two copies of GR.
In the case of the two copies of GR, after deriving all the equations of motion, with use of the two copies of the four-dimensional diffeomorphism invariance, for the perturbed metrics \eqref{metric_g_ell0} and \eqref{metric_f_ell0} we may choose the gauge 
\begin{eqnarray}
h_1=0,\qquad k_1=0,
\qquad
n_2=0,\qquad m_2=0.
\end{eqnarray}
For convenience, we introduce the new variables $\psi$ and $\chi$
\begin{eqnarray}
n_1
=
n_0
+
\frac{1}{4\sqrt{r_g}}
\psi,
\qquad
m_1
=
m_0
+
\frac{1}{4\sqrt{r_f}}
\chi,
\end{eqnarray}
Integrating the equations of motion for $n_0$ and $m_0$, we find the solutions
\begin{eqnarray}
\label{psichi}
\psi
=
C_\psi (t),
\qquad 
\chi
=
C_\chi (t),
\end{eqnarray}
where $C_\psi(t)$ and $C_\chi (t)$ are functions of time.
The equations of motion for $n_2$ and $m_2$ then, respectively, lead to
\begin{eqnarray}
\label{c_constraint}
C_\psi (t)
=
C_{\psi,0},
\qquad 
C_\chi (t)
=
C_{\chi,0},
\end{eqnarray}
where $C_{\psi,0}$ and $C_{\chi,0}$ are  integration constants.
The equations of motion for $\psi$ and $\chi$, respectively, give rise to
\begin{eqnarray}
n_0=C_{n_0}(t),
\qquad 
m_0=C_{m_0}(t),
\end{eqnarray}
where $C_{n_0}(t)$ and $C_{m_0} (t)$ are free functions of time.
We note that the constants $C_{\psi,0}$ and $C_{\chi,0}$ can be absorbed into the redefinition of gravitational radii $r_g$ and $r_f$, and the functions $C_{n_0}(t)$ and $C_{m_0} (t)$ can be set to zero by the redefinition of the time coordinates in each sector.
Thus, the $\ell=0$ mode in the case of the two copies of GR can be absorbed by the redefinition of the Schwarzschild backgrounds.

\subsection{The case of MTBG}

We then focus on the $\ell =0$ mode in the self-accelerating branch of MTBG.
For convenience, we also introduce the new variables $\psi$ and $\chi$
\begin{eqnarray}
\label{n_1}
&&
n_1
=
n_0
+
\frac{1}{4r_g}
\Big[
  (6r-5r_g)h_1
-2 r n_2
+2r(r-r_g) h_1'
+
2\sqrt{r^3r_g}
\dot{h}_1
+
\sqrt{r_g}
\psi
\Big],
\\
\label{m_1}
&&
m_1
=
m_0
+
\frac{1}{4r_f}
\Big[
  (6b^2r-5r_f)k_1
-2b^2 r m_2
+2b^2r^2 k_1'
-2rr_f k_1'
+
2\sqrt{r^3r_f}
\dot{k}_1
+
\sqrt{r_f}
\chi
\Big].
\end{eqnarray}
Integrating the equations of motion for $n_0$ and $m_0$, respectively, we find the solutions
\begin{eqnarray}
\label{psichi}
\psi
=
C_\psi (t)
+
\frac{2(r_g-2 r)}
        {\sqrt{r_g}}
h_1,
\qquad 
\chi
=
C_\chi (t)
+
\frac{2(r_f-2b^2 r)}
        {\sqrt{r_f}}
k_1,
\end{eqnarray}
where $C_\psi(t)$ and $C_\chi (t)$ are functions of time.
The equations of motion for $\lambda_0$ and $\lambda_1$ yield
\begin{eqnarray}
\label{m_2}
m_2
&=&
\frac{1}{2}
\left(
h_1-
k_1
\right)
+
n_2.
\end{eqnarray}
The equation of motion for $\lambda_2$ can then be integrated as 
\begin{eqnarray}
\label{k_1}
k_1
=
h_1
+
\frac{C_{hk} (t)}
        {r^{\frac{3}{2}}},
\label{eq_lam0_0}
\end{eqnarray}
where $C_{hk} (t)$ is a function of time.
The equation of motion for $m_2$  leads to 
\begin{eqnarray}
\lambda_2
&
=&\frac{1}
          {2bC_0^2 (C_0c_1+c_2) (-1+C_0\beta)m^2r^{\frac{3}{2}}}
\Big[
-(b-1)bC_0^3 (C_0c_1+c_2)m^2r C_{hk}(t)
+2C_0^2 (C_0c_1+c_2)m^2 r
\left(\sqrt{r_f}-b\sqrt{r_g}\right)
\lambda_0
\nonumber
\\
&&
+
C_0^3r\alpha^2 
C_\chi'(t)
-
4C_0 c_1 m^2\sqrt{r_f}
\lambda_1'
-
4c_2 m^2\sqrt{r_f}
\lambda_1'
-
4bC_0^3c_1 m^2\sqrt{r_g}
\lambda_1'
-
4bC_0^2c_2 m^2\sqrt{r_g}
\lambda_1'
\nonumber\\
&&
-
2C_0c_1 m^2 
r\sqrt{r_f}
\lambda_1''
-
2c_2m^2 r\sqrt{r_f}
\lambda_1''
-
2bC_0^3c_1m^2r\sqrt{r_g}
\lambda_1''
-
2bC_0^2c_2 m^2r\sqrt{r_g}
\lambda_1''
\Big].
\end{eqnarray}
Similarly, the equation of motion for $n_2$ leads to 
\begin{eqnarray}
\label{c_constraint}
C_\psi (t)
=
C_{\psi,0}
-
\frac{C_0^2\alpha^2}{b}
C_{\chi}(t),
\end{eqnarray}
where $C_{\psi,0}$ is an integration constant.

The equation of motion for $\psi$ reduces to the equation
\begin{eqnarray}
\label{comb1}
2h_1'-2n_0'-n_2'+rh_1''
+
\sqrt{\frac{r}{r_g}}
\left(
-\dot{h}_1+\dot{n}_2
-
r
\dot{h}_1'
\right)
=0,
\end{eqnarray}
and the equation of motion for $\chi$ reduces to the equation
\begin{eqnarray}
\label{comb2}
2h_1'-2m_0'-n_2'+rh_1''
+
\sqrt{\frac{r}{r_f}}
\left(
-\dot{h}_1+\dot{n}_2
-r \dot{h}_1'
\right)
=0.
\end{eqnarray}
The compatibility of the equations of motion for $\psi$ and $h_1$ leads to the solution
\begin{eqnarray}
\label{s_l0}
\lambda_0
=
C_{\lambda_0}(t)
+
\frac{\sqrt{r_f}+bC_0^2\sqrt{r_g}}
        {C_0^2(\sqrt{r_f}-b\sqrt{r_g})}
\left(
\lambda_1''
+
\frac{2}{r}
\lambda_1'
\right),
\end{eqnarray}
where $C_{\lambda_0}(t)$ is a free function of time.
The compatibility of the equations of motion for $\chi$ and $k_1$ leads to the same solution as Eq.~\eqref{s_l0}.
The combination of Eqs.~\eqref{comb1} and \eqref{comb2} can be integrated as 
\begin{eqnarray}
\label{con}
m_0
=
C_{m_0}(t)
+
\sqrt{\frac{r_g}{r_f}}
n_0
+
\frac{\sqrt{r_f}-\sqrt{r_g}}
        {2\sqrt{r_f}}
\left(
h_1
+rh_1'
-n_2
\right),
\end{eqnarray}
where $C_{m_0}(t)$ is a free function of time.
Then, Eq.~\eqref{comb2} reduces to Eq.~\eqref{comb1}.

So far, we have not fixed the gauge.
From now on, we fix the gauge as Eq.~\eqref{gauge_ell0} and then Eq.~\eqref{comb1} reduces to
\begin{eqnarray}
\label{n0n2}
-2n_0'-n_2'
+
\sqrt{\frac{r}{r_g}}
\dot{n}_2
=0.
\end{eqnarray}
Under the condition \eqref{n0n2}, the general solution is given by 
\begin{eqnarray}
n_1
&=&
\frac{C_{\psi_0}}{4\sqrt{r_g}}
-
\frac{C_0^2\alpha^2}{4b\sqrt{r_g}}
C_\chi (t)
+
n_0
-
\frac{r}{2r_g}
n_2,
\qquad 
m_0
=
C_{m_0}(t)
+
\sqrt{\frac{r_g}{r_f}}
n_0
-
\frac{\sqrt{r_f}-\sqrt{r_g}}
        {2\sqrt{r_f}}
n_2,
\nonumber
\\
m_1
&=&
C_{m_0}(t)
+
\frac{1}{4\sqrt{r_f}}
C_{\chi}(t)
+
\frac{1}
       {2\sqrt{r_f}}
C_{hk}'(t)
+
\frac{1}{2r_f}
\left[
2\sqrt{r_f r_g}n_0
+
\left(
-b^2r-r_f+\sqrt{r_fr_g}
\right)
n_2
\right],
\nonumber
\\
m_2
&=&
n_2
-
\frac{1}
       {2r^{\frac{3}{2}}}
C_{hk}(t),
\qquad 
k_1
=
\frac{1}
       {r^{\frac{3}{2}}}
C_{hk}(t).
\end{eqnarray}
Imposing for simplicity the flatness of the spatial three-dimensional part of the $g_{\mu\nu}$ metric, we have $n_2=0$ and then the integration of Eq.~\eqref{n0n2} is given by $n_0=C_{n_0}(t)$, where $C_{n_0}(t)$ is a free function of time.
Imposing for simplicity the flatness of the spatial three-dimensional part of the  $f_{\mu\nu}$ metric, we have $m_2=k_1=0$, and hence $C_{hk}(t)=0$.
In general, imposing spatial flatness in the two sectors restricts the solution space for the $\ell=0$ perturbations.
However, we nonetheless choose to impose this condition to be compatible with the discussion in subsection \ref{sec4c}.
By the reparametrization,
\begin{eqnarray}
\label{reparametrization}
C_{m_0}(t)
=
d_{m_0}(t)
+
\left(
1-\sqrt{\frac{r_g}{r_f}}
\right)
C_{n_0}(t),
\end{eqnarray}
where $d_{m_0}(t)$ is a function of time, 
the metric solution with the flatness of the three-dimensional spatial metrics
is given by 
\begin{eqnarray}
\label{metric_solution_ell0}
n_0
&=&
C_{n_0}(t),
\qquad 
n_1
=
C_{n_0}(t)
+
\frac{1}{4\sqrt{r_g}}
\left(
C_{\psi_0}
-
\frac{C_0^2\alpha^2}{b}
C_\chi (t)
\right),
\qquad 
n_2
=
0,
\nonumber
\\
m_0
&=&
C_{n_0}(t)
+
d_{m_0}(t),
\qquad
m_1
=
C_{n_0}(t)
+
d_{m_0}(t)
+
\frac{1}{4\sqrt{r_f}}
C_{\chi}(t),
\qquad
m_2
=
k_1
=0,
\end{eqnarray}
which satisfies the constraint
\begin{eqnarray}
\label{linear_constraint}
\sqrt{r_g}
\left(
n_1-n_0
\right)
+
\frac{C_0^2\alpha^2\sqrt{r_f}}
       {b}
(m_1-m_0)
=
\frac{C_{\psi_0}}
       {4}.
\end{eqnarray}
$C_{n_0}(t)$ corresponds to the DOF of the redefinition of the time coordinate $t$.
In other words, we may set $C_{n_0}(t)=0$ by a suitable redefinition of the time coordinate.
Although the metric solution satisfies the flatness of the spatial three-dimensional metrics and the asymptotic flatness of  the two spacetimes, there are still the two functions of time, $C_{\chi}(t)$ and $d_{m_0}(t)$.
We also impose that the two sectors share the same asymptotic Minkowski vacua, and set $d_{m_0}(t)=0$.

The general solution for the Lagrange multipliers which do not affect the metric solutions is explicitly given by 
\begin{eqnarray}
\label{sol_lambda0}
\lambda_0
&=&
C_{\lambda_0}(t)
+
\frac{\sqrt{r_f}+bC_0^2\sqrt{r_g}}
        {C_0^2\left(\sqrt{r_f}-b\sqrt{r_g}\right)}
\left(
\lambda_1''
+
\frac{2}{r}
\lambda_1'
\right),
\\
\lambda_2
&=&
\frac{2(C_0c_1+c_2)m^2 \left(\sqrt{r_f}-b\sqrt{r_g}\right)C_{\lambda_0}(t)+C_0\alpha^2C_\chi'(t)}
        {2b(C_0c_1+c_2)m^2\sqrt{r}(-1+C_0\beta)},
\label{sol_lambda2}
\end{eqnarray}
while $\lambda_1$ is undetermined.
If we consider the gravitational collapse of spherically symmetric stars in both sectors, then one should impose the regularity at the center and it is expected that the function $C_{\chi}(t)$ should be fixed to a constant value that results in a constant mass of matter in the interior region of each sector,
\begin{eqnarray}
C_{\chi}(t)=C_{\chi,0}={\rm const}.
\end{eqnarray}
Thus, the  solutions for the Lagrange multipliers reduce to 
\begin{eqnarray}
\lambda_0
&=&
C_{\lambda_0}(t)
+
\frac{\sqrt{r_f}+bC_0^2\sqrt{r_g}}
        {C_0^2\left(\sqrt{r_f}-b\sqrt{r_g}\right)}
\left(
\lambda_1''
+
\frac{2}{r}
\lambda_1'
\right),
\qquad 
\lambda_2
=
\frac{ \left(\sqrt{r_f}-b\sqrt{r_g}\right)C_{\lambda_0}(t)}
        {b\sqrt{r}(-1+C_0\beta)}.
\label{sol_lambda2}
\end{eqnarray}
Because of the regularity of $\lambda_2$ as $r\to \infty$, no condition is imposed on $C_{\lambda_0}(t)$.
On the other hand, $\lambda_0$ and $\lambda_1$ are not determined by the equations of motion for the $\ell=0$ perturbations.
\footnote{At higher-order the metric may depend on $\lambda_0$ and/or $\lambda_1$. If this is the case then $\lambda_0$ and/or $\lambda_1$ may be fixed by suitable boundary conditions for higher-order perturbations. However, this is beyond the scope of the present paper.}

The manipulations to reduce the equations of motion for the $\ell=0$ mode are summarized in Table~\ref{tablel0}.
In the next subsection, we confirm that the general solution for the $\ell=0$ mode corresponds to the linearized limit of the time-dependent extension of the Schwarzschild solution in the spatially-flat coordinates in the self-accelerating branch of MTBG.

\begin{table}[htbp]
\begin{center}
\begin{tabular}{ |c|c|c|c| } 
\hline
 Manipulation & Output & Remark & \# of  independent variables \\
\hline
\hline
Derive EOMs &  & & 11\\ 
EOM for $n_0$ & Eliminate $\psi$ & $C_{\psi}(t)$ & 10\\
EOM for $m_0$ & Eliminate $\chi$ & $C_{\chi}(t)$ & 9\\ 
EOM for $\lambda_0$  & Eliminate $m_2$&  EOM for $\lambda_1$ is not independent & 8\\ 
EOM for $\lambda_2$  & Eliminate $k_1$ & $C_{hk}(t)$ &7 \\
EOM for $m_2$  & Eliminate $\lambda_2$ & &6 \\
EOM for $n_2$  & & Eliminate $C_{\psi}(t)$ &6 \\
Combine EOM for $\psi$ and EOM for $h_1$  & Eliminate $\lambda_0$ & $C_{\lambda_0}$ & 5\\
Combine EOM for $\psi$ and EOM for $\chi$  & Eliminate $m_0$ & $C_{m_0}$ & 4 \\
Fix gauge & Eliminate $h_1$ &  &3 \\
Spatial flatness of $g_{\mu\nu}$  and $f_{\mu\nu}$
& Set $n_2 = 0$  & Eliminate $C_{hk}(t)$& 2 \\
EOM for $\psi$ & Eliminate $n_0$ & $C_{n_0} (t)$ &1 \\
Redefinition of time & & Eliminate $C_{n_0} (t)$ &1\\
Same Minowski vacua &  & Eliminate $C_{m_0} (t)$ & 1\\
Argument on collapse  &
& Eliminate $C_\chi(t)$ & 0 \\
\hline
\end{tabular}
\caption{
The manipulations to reduce the equations of motion for the $\ell=0$ mode.
}
\label{tablel0}
\end{center}
\end{table}

\subsection{Time-dependent extensions of the Schwarzschild solutions in the self-accelerating branch of MTBG}
\label{sec4c}

The most general spherically symmetric but time-dependent physical and fiducial metrics are given by, respectively, 
\begin{eqnarray}
\label{generalfg}
g_{\mu\nu}
dx^\mu dx^\nu
&=&
-A_0(t,r)dt^2
+A_1(t,r)
\left(dr+N^r(t,r) dt\right)^2
+A_2(t,r)r^2 
\left(
 d\theta^2
+\sin^2\theta d\varphi^2
\right),
\nonumber\\
f_{\mu\nu}
dx^\mu dx^\nu
&=&
-A_{0f}(t,r)dt^2
+A_{1f}(t,r)\left(dr^2+N^r_f(t,r) dt\right)^2
+ A_{2f}(t,r)
r^2 
\left(
 d\theta^2
+\sin^2\theta d\varphi^2
\right).
\end{eqnarray}
Under the spherical symmetry, the most general ansatz for the Lagrange multipliers is given by Eq.~\eqref{spherical_lagrange}.

Within the general ansatz of the metrics \eqref{generalfg}, we assume the ansatz for the time-dependent Schwarzschild metrics in the spatially-flat coordinates
\begin{eqnarray}
\label{gp2}
&&
A_0=C_n(t)^2,
\qquad
A_1=A_2=1,
\qquad 
N^r
=
C_g(t)
\sqrt{
\frac{1}{r}},
\nonumber\\
&&
A_{0f}=b^2C_0^2 C_{m}(t)^2,
\qquad 
A_{1f}=A_{2f}=C_0^2,
\qquad 
N^r_f
=
C_f(t)
\sqrt{
\frac{1}{r}}.
\end{eqnarray}
We also impose the condition for the self-accelerating branch Eq.~\eqref{sa} and the vanishing effective cosmological constants \eqref{flat}, under which the equations of motion for $\lambda$, ${\bar \lambda}$, and $\lambda^r$ are automatically satisfied. The equations of motion for $A_0$, $N^r$, $A_{0f}$, and $N^r_f$ are also automatically satisfied.

The equation of motion for $A_1$ relates $\lambda^r$ with other variables as
\begin{eqnarray}
{\lambda}^r
&=&
\frac{1}{\sqrt{r}}
\frac{bC_0^2 C_g(t)C_m(t)+C_f(t) C_n(t)}
        {bC_0^2C_m(t)C_n (t)\left(1-C_0\beta\right)}
\left(
{\bar \lambda}''
+
\frac{2}{r}
{\bar \lambda}'
\right)
+
\frac{1}{\sqrt{r}}
\frac{-bC_g(t)C_m(t)+C_f(t)C_n(t)}
        {b C_m(t)C_n(t)\left(-1+C_0\beta\right)}
\lambda
\nonumber
\\
&-&
\frac{2}{\sqrt{r}}
\frac{C_n(t)C_g'(t)-C_g(t)C_n'(t)}
       {C_0 (C_0c_1+c_2)m^2(-1+C_0\beta) C_n(t)^2}.
\end{eqnarray}
The equation of motion for $A_2$ yields
\begin{eqnarray}
\lambda
=
C_\lambda(t)
-
\frac{bC_0^2C_g(t) C_m(t)+C_f(t)C_n (t)}
        {C_0^2 \left(bC_g(t)C_m(t)-C_f(t)C_n(t)\right)}
\left(
{\bar \lambda}''
+
\frac{2}{r}
{\bar \lambda}'
\right).
\end{eqnarray}
The equations of motion for $A_{1f}$ and $A_{2f}$ provide a degenerate equation, which can be integrated as 
\begin{eqnarray}
\label{cg}
C_g(t)
=
C_n(t)
\left(
C_{g,0}
-
\frac{C_0^2\alpha^2}{b}
\frac{C_f(t)}{C_m(t)}
\right),
\end{eqnarray}
where $C_{g,0}$ is an integration constant. Then, all the components of the equations of motion are satisfied.

By the redefinition of the functions of time, $C_m(t)=C_n(t)d_m(t)$ and $C_f(t) =C_n(t) d_f(t)d_m(t)$, where $d_m(t)$ and $d_f(t)$ are functions of time, from Eq.~\eqref{cg} we obtain
$C_g(t)=C_n(t)
\left(
C_{g,0}
-
\frac{C_0^2\alpha^2}{b}
d_f(t)
\right)$.
The nontrivial metric solution is given by 
\begin{eqnarray}
\label{nonlinear_metric}
&&
A_0=C_n(t)^2,
\qquad 
N^r=\sqrt{\frac{1}{r}} 
\left(
C_{g,0}
-
\frac{C_0^2\alpha^2}{b}
d_f(t)
\right)
C_n(t),
\nonumber\\
&&
A_{0f}=b^2C_0^2 d_m(t)^2C_n(t)^2,
\qquad
N^r_f=\sqrt{\frac{1}{r}}
d_f(t)d_m (t)C_n(t),
\end{eqnarray}
which satisfies
\begin{eqnarray}
\sqrt{r}
\left(
\frac{N^r}{C_n(t)}
+
\frac{C_0^2\alpha^2}{b}
\frac{M^r}{C_m(t)}
\right)
=
C_{g,0},
\end{eqnarray}
corresponding to the nonlinear extension of Eq.~\eqref{linear_constraint}.
While ${\bar \lambda}$ is undetermined by the equations of motion,
the general solutions for the Lagrange multipliers $\lambda$ and $\lambda^r$ are given by 
\begin{eqnarray}
\label{sol_lambda}
{\lambda}
&=&
C_\lambda(t)
-
\frac{bC_0^2C_{g,0}+d_f(t)-C_0^4\alpha^2 d_f(t)}
        {C_0^2 \left(b C_{g,0}-d_f(t)-C_0^2\alpha^2d_f(t)\right)}
\left(
{\bar \lambda}''
+
\frac{2}{r}
{\bar \lambda}'
\right),
\\
\label{sol_lambda_r}
\lambda^r
&=&
\frac{1}{\sqrt{r}}
\frac{(C_0c_1+c_2)m^2C_\lambda(t)\left(-bC_{g,0}+d_f(t)+C_0^2\alpha^2 d_f(t)\right)+2C_0\alpha^2 d_f'(t)}
       {b m^2(C_0c_1+c_2)(-1+C_0\beta)}.
\end{eqnarray}
By the rescaling of the time coordinate, we may set $C_{n}(t)=1$.
In the absence of BHs, the solution reduces to the Minkowski solutions.
In order for both the sectors to share the same asymptotic Minkowski vacua as $r\to \infty$, we impose $d_m(t)=1$ and then obtain
\begin{eqnarray}
\label{reduced_spherical_metric_solution}
&&
A_0=1,
\qquad 
N^r=\sqrt{\frac{1}{r}} 
\left(
C_{g,0}
-
\frac{C_0^2\alpha^2}{b}
d_f(t)
\right),
\qquad 
A_{0f}=b^2C_0^2,
\qquad
N^r_f=\sqrt{\frac{1}{r}}
d_f(t).
\end{eqnarray}
If we consider spherically symmetric stellar solutions in both the sectors, Eq.~\eqref{reduced_spherical_metric_solution} would describe the exterior solution outside the stars.
Imposing the regularity at the center and integrating the corresponding equation of motion towards the exterior, the coefficient of the $1/\sqrt{r}$ terms in $N^r$ and $N^r_f$ should be determined by the total masses in the interior of the stars.
Thus, $d_f(t)$ should be fixed to a constant value, because the total mass of matter forming the star in each sector has to be constant,
\begin{eqnarray}
d_f(t)=d_{f,0}.
\end{eqnarray}
We note that ${\bar \lambda}$ is not fixed by the equations of motion, and 
the general solution for $\lambda$ and $\lambda^r$ is then given by
\begin{eqnarray}
\lambda
&=&
C_{\lambda}(t)
+
\frac{bC_0^2C_{g,0}+d_{f,0}-C_0^4d_{f,0}\alpha^2}
        {C_0^2\left(-bC_{g,0}+d_{f,0} +C_0^2\alpha^2 d_{f,0}\right)}
\left(
{\bar\lambda}''
+
\frac{2}{r}
{\bar\lambda}'
\right),
\\
\lambda^r
&=&
\frac{bC_{g,0}-d_{f,0}\left(1+C_0^2\alpha^2\right)}
        {b\left(1-C_0\beta\right)\sqrt{r}}
C_{\lambda}(t).
\end{eqnarray}
Because of the regularity of $\lambda_r$ as $r\to \infty$, no condition is imposed for $C_{\lambda}(t)$.
$\lambda$ and ${\bar \lambda}$ are not determined by the equations of motion in the spherically symmetric backgrounds.
\footnote{Introducing small deviations from the spherically symmetry, the behavior of the metric may depend on $\lambda$ and/or ${\bar \lambda}$. If this is the case then $\lambda$ and/or ${\bar \lambda}$ may be fixed by some boundary conditions for the small deviations from the spherically symmetry. However,  this is beyond the scope of the present paper.
}

\section{The solutions of the dipolar perturbations}
\label{sec5}

In this section, we focus on the $\ell = 1$ mode.

\subsection{The case of the two copies of GR}

In this subsection, we review the case of the two copies of GR, to help illustrate the more complex case of MTBG.
In the perturbed metrics for the $\ell=1$  mode, Eqs.~\eqref{metric_g} and \eqref{metric_f}, 
we set $h_2=0$ and $k_2=0$, and under the two copies of the four-dimensional diffeomorphism invariance we may choose the gauge 
\begin{eqnarray}
h_t=0,
\qquad 
h_r=0,
\qquad
h_1=0,
\qquad 
k_t=0,
\qquad 
k_r=0,
\qquad
k_1=0.
\end{eqnarray}
Similar to the case of $\ell=0$, we introduce the new variables $\psi$ and $\chi$ to eliminate the perturbations of the radial components of the shift vectors $n_1$ and $m_1$
\begin{eqnarray}
n_1
=
\frac{1}{2r r_g}
\Big(
\sqrt{rr_g}\psi
+
2rr_g n_0
-
r^2 n_2
\Big),
\qquad 
m_1
=
\frac{1}{2r r_f}
\Big(
\sqrt{rr_f}\chi
+
2rr_f m_0
-
b^2 r^2 m_2
\Big).
\end{eqnarray}
Using the equations of motion for $n_0$ and $m_0$, we can relate $n_0$ and $m_0$ with other variables as 
\begin{eqnarray}
n_0
=
-\frac{1}{2r_g}
\Big(
rn_2
+2\sqrt{rr_g}\psi'
\Big),
\qquad
m_0
=
-
\frac{1}{2r_f}
\Big(
b^2rm_2
+2\sqrt{rr_f}\chi'
\Big).
\end{eqnarray}
The equations of motion for $\psi$ and $\delta h$ yield, respectively, the evolution equations of $n_2$. 
Their degeneracy leads to 
\begin{eqnarray}
n_2=\frac{1}{3}\sqrt{\frac{r}{r_g}}\left(3\psi'+2r\psi''\right).
\end{eqnarray}
Substituting it into the equation of motion for $\psi$ or the equivalent equation of motion for $\delta h$,
\begin{eqnarray}
\label{ell1psi}
2r(r-r_g)\psi'+2\sqrt{r^3r_g}\dot{\psi}+(r+r_g)\psi=0,
\end{eqnarray}
with which all the other equations of motion in the $g_{\mu\nu}$ sector can be satisfied.
Similarly, the equations of motion for $\chi$ and $\delta k$ yield, respectively, the evolution equations of $m_2$, whose degeneracy leads to  
\begin{eqnarray}
m_2=\frac{1}{3}\sqrt{\frac{r}{r_f}}\left(3\chi'+2r\chi''\right).
\end{eqnarray}
Substituting it into the equation of motion for $\psi$ or the equivalent equation of motion for $\delta k$, 
\begin{eqnarray}
\label{ell1chi}
2r(b^2r-r_f)\chi'+2\sqrt{r^3r_f}\dot{\chi}+(b^2r+r_f)\chi=0,
\end{eqnarray}
with which all the other equations of motion in the $f_{\mu\nu}$ sector can be satisfied.
Since the master equations \eqref{ell1psi} and \eqref{ell1chi} are of the first order, there is no propagating DOF.
In order to satisfy the regularity boundary conditions at spatial infinity and at the horizon, we obtain the trivial solution
\begin{eqnarray}
\psi=0,\qquad \chi=0,
\end{eqnarray}
as the solutions of the $\ell=1$ mode in the case of the two copies of GR.

\subsection{The case of MTBG}

We then focus on the self-accelerating branch of  MTBG.
In order to make the analysis of the $\ell=1$ mode more explicit, we focus on the case of the effective massless case with $b=1$ and $r_f=r_g$.
Even with these conditions, the constraint part of the Lagrangian \eqref{lag22} does not vanish and hence the essential properties of MTBG are still retained.

After deriving the equations of motion, 
similar to the case of $\ell=0$, we introduce the new variables $\psi$ and $\chi$ to eliminate the perturbations of the radial components of the shift vectors, $n_1$ and $m_1$
\begin{eqnarray}
n_1
&=&
\frac{1}{4r r_g}
\Big[
2\sqrt{rr_g}\psi
+
r
\left(
6r-5r_g
\right)
h_1
+
4r(r-r_g)
h_r
+
4rr_g n_0
\nonumber\\
&&
-
2r^2n_2
+
2r^2
(r-r_g)
h_1'
+
2\sqrt{r^5r_g}
\left(
\delta h
+
\dot{h}_1
\right)
\Big].
\nonumber\\
m_1
&=&
\frac{1}{4r r_g}
\Big[
2\sqrt{rr_g}\chi
+
r
\left(
6r-5r_g
\right)
k_1
+
4r(r-r_g)
k_r
+
4rr_g m_0
\nonumber\\
&&
-
2r^2m_2
+
2r^2
(r-r_g)
k_1'
+
2\sqrt{r^5r_g}
\left(
\delta k
+
\dot{k}_1
\right)
\Big].
\end{eqnarray}
We also  replace $h_t$ and $k_t$ by $\delta h$ and $\delta k$, respectively, by
\begin{eqnarray}
\label{deltahk}
{h}_t
=
-
\sqrt{\frac{r_g}{r^3}}
{h}_r
+
\delta {h},
\qquad 
{k}_t
=
-
\sqrt{\frac{r_g}{r^3}}
{k}_r
+
\delta {k}.
\end{eqnarray}
In the case of the two copies of GR, with use of the temporal gauge DOFs, we can set $\delta h=0$ and $\delta k=0$.
In the case of MTBG, since there is no temporal gauge DOF, we cannot set $\delta h=0$ and $\delta k=0$ and instead treat $\delta h$ and $\delta k$ as independent variables.
We then derive the equations of motion for the metric perturbations
$n_0$, $\psi$, $n_2$, $\delta h$, $m_0$, $\chi$, $m_2$, $\delta k$, $k_r$, and $k_1$,
and also those for the perturbations of the Lagrange multipliers $\lambda_0$, $\lambda_1$, $\lambda_2$, and $\lambda_3$.

Using the equations of motion for $n_0$ and $m_0$, we can relate $n_0$ and $m_0$ with the other variables as 
\begin{eqnarray}
n_0
&=&
\frac{1}{4r_g}
\Big\{
4r^{\frac{3}{2}}\sqrt{r_g}\delta h
+
(-18r+5r_g)h_1
\nonumber\\
&&
-
2
\Big[
-2(r-2r_g)h_r
+rn_2
+5r^2 h_1'
-3rr_g h_1'
+2\sqrt{rr_g}\psi'
+r^{\frac{3}{2}}\sqrt{r_g} \dot{h}_1
-2r^{\frac{3}{2}}\sqrt{r_g} \dot{h}_r
\Big]
\Big\},
\nonumber\\
m_0
&=&
\frac{1}{4r_g}
\Big\{
4r^{\frac{3}{2}}\sqrt{r_g}\delta k
+
(-18r+5r_g)k_1
\nonumber\\
&&
-
2
\Big[
-2(r-2r_g)k_r
+rm_2
+5r^2 k_1'
-3rr_g k_1'
+2\sqrt{rr_g}\chi'
+r^{\frac{3}{2}}\sqrt{r_g} \dot{k}_1
-2r^{\frac{3}{2}}\sqrt{r_g} \dot{k}_r
\Big]
\Big\}.
\nonumber
\end{eqnarray}
In the effectively massless case $b=1$ and $r_f=r_g$, the equation of motion for $\lambda_0$ becomes trivial.
The equation of motion for $\lambda_1$
yields
\begin{eqnarray}
m_2
=
n_2
+
\frac{1}{2}
\left(
h_1
-
k_1
\right).
\end{eqnarray}
The equations for  $\lambda_2$ and $\lambda_3$ can be integrated as
\begin{eqnarray}
k_r=
h_r+ \frac{1}{r^2}C_{k_1}(t)+\frac{1}{r^{\frac{5}{2}}} C_{k_r}(t),
\qquad 
k_1
=
h_1+
\frac{2}{r^2}C_{k_1}(t)+\frac{1}{r^{\frac{5}{2}}} C_{k_r}(t),
\end{eqnarray}
where $C_{k_1}(t)$ and $C_{k_r}(t)$ are free functions of time.
The equations of motion for $\psi$ and $\delta h$ yield, respectively, the evolution equations of $n_2$ as
\begin{eqnarray}
\label{psih}
\dot{n}_2=F_{n_2,1},
\qquad 
\dot{n}_2=F_{n_2,2},
\end{eqnarray}
where
\begin{eqnarray}
F_{n_2,1}
&:=&
\frac{7}{2}
\delta h
-
\frac{1}{2r^2}
\psi
+
3r
\delta h'
-
\sqrt{\frac{r}{r_g}}
n_2'
+
\sqrt{\frac{r_g}{r}}
n_2'
-
2\psi''
\nonumber\\
&-&
\frac{6}{\sqrt{rr_g}}h_1
+\sqrt{\frac{r_g}{r^3}}h_r
-17\sqrt{\frac{r}{r_g}}h_1'
+\frac{5}{2}\sqrt{\frac{r_g}{r}}h_1'
+2\sqrt{\frac{r}{r_g}}h_r'
-6\sqrt{\frac{r_g}{r}}
        h_r'
-\frac{5r^{\frac{3}{2}}}{\sqrt{r_g}}h_1''
\nonumber\\
&+&2\sqrt{rr_g}h_1''
-
\frac{1}{2}\dot{h}_1
+3 \dot{h}_r
+2r \dot{h}_r',
\nonumber
\\
F_{n_2,2}
&:=&
\frac{1}{2}
\Big(
-
13\delta h
-
6\sqrt{\frac{r_g}{r^3}}
n_2
-
\frac{\psi}{r^2}
-
16 r\delta h'
-
2\sqrt{\frac{r}{r_g}}
n_2'
+
2\sqrt{\frac{r_g}{r}}
n_2'
+
\frac{6}{r}
\psi'
-
4r^2\delta h''
\nonumber\\
&+&
\frac{(24r-5r_g)}{\sqrt{r^3r_g}}
h_1
+
20
\sqrt{\frac{r_g}{r^3}}
h_r
+
20
\sqrt{\frac{r}{r_g}}
h_1'
-5
\sqrt{\frac{r_g}{r}}
h_1'
+
4
\sqrt{\frac{r}{r_g}}
h_r'
+
\frac{2r^{\frac{3}{2}}}{\sqrt{r_g}}
h_1''
-
\dot{h}_1
+
6\dot{h}_r
+
4r \dot{h}_r'
\Big).
\end{eqnarray}
Similarly, the equations of motion for $\chi$ and $\delta k$ yield, respectively, the other  evolution equations of $n_2$ as 
\begin{eqnarray}
\label{chik}
\dot{n}_2=G_{n_2,1},
\qquad 
\dot{n}_2=G_{n_2,2},
\end{eqnarray}
where
\begin{eqnarray}
G_{n_2,1}
&:=&
\frac{1}{2r^{9/2}r_g}
\Big[
2r\sqrt{r_g} (-10r+29r_g)C_{k_1} (t)
+
3\sqrt{rr_g} (-9r+19r_g)C_{k_r}(t)
-
r^2\sqrt{r_g}
\Big(
-7\sqrt{r^5r_g}
\delta k
\nonumber\\
&&
+
\sqrt{rr_g}\chi
+
2\sqrt{rr_g}
C_{k_1}'(t)
+
4\sqrt{r_g}
C_{k_r}'(t)
-
6
\sqrt{r^7r_g}
\delta k'
+
2r^3 n_2'
-
2r^2r_g 
n_2'
+
4\sqrt{r^5r_g}\chi''
\nonumber
\\
&&
+12 r^2h_1
-2rr_g h_r
+34 r^3 h_1'
-5r^2r_g h_1'
-4r^3 h_r'
+12r^2r_gh_r'
+10r^4 h_1''
\nonumber\\
&&
-4r^3r_g h_1''
+
\sqrt{r^5r_g}
\dot{h}_1
-6\sqrt{r^5r_g}
\dot{h}_r
-4\sqrt{r^7r_g} 
 \dot{h}_r'
\Big)
\Big],
\nonumber
\\
G_{n_2,2}
&:=&
\frac{1}{2r^4\sqrt{r_g}}
\Big[
-
20\sqrt{r} (r-2r_g)C_{k_1}(t)
+
(-21r+33r_g)C_{k_r}(t)
-
13r^4 \sqrt{r_g}
\delta k
-
6
r^{\frac{5}{2}}r_g
n_2
-
r^2\sqrt{r_g}
\chi
\nonumber\\
&&
-
2r^2\sqrt{r_g}
C_{k_1}'(t)
-
4\sqrt{r^3r_g}
C_{k_r}'(t)
-
16r^5\sqrt{r_g}
\delta k'
-
2r^{\frac{9}{2}}
n_2'
+
2r^{\frac{7}{2}}r_g
n_2'
+
6r^3 \sqrt{r_g}
\chi'
-
4r^6\sqrt{r_g}
\delta k''
\nonumber\\
&&
+24r^{\frac{7}{2}}h_1
-5 r^{\frac{5}{2}}r_g h_1
+20 r^{\frac{5}{2}}r_g h_r
+20 r^{\frac{9}{2}}h_1'
-5r^{\frac{7}{2}}r_g h_1'
+4r^{\frac{9}{2}}h_r'
+2r^{\frac{11}{2}} h_1''
-r^4\sqrt{r_g} \dot{h}_1
\nonumber
\\
&&
+6r^4\sqrt{r_g} \dot{h}_r
+4r^5 \sqrt{r_g}\dot{h}_r'
\Big].
\end{eqnarray}
The consistency of Eq.~\eqref{psih}, $F_{n_2,1}=F_{n_2,2}$, yields
\begin{eqnarray}
\label{sol_n2_l1}
n_2
&=&
\frac{1}{3}
\sqrt{\frac{r}{r_g}}
\Big(
-
10r \delta h
-
11r^2\delta h'
+
3\psi'
-
2r^3\delta h''
+
2r\psi''
\Big)
\nonumber \\
&&
+
\frac{1}{6r_g}
\Big(
\big(36r-5r_g\big)h_1
+
2
\big(
9r_g h_r
+
27r^2h_1'
-
5rr_g h_1'
+
6rr_g h_r'
+
6r^3h_1''
-
2r^2r_g h_1''
\big)
\Big).
\end{eqnarray}
The consistency of Eq.~\eqref{chik}, $G_{n_2,1}=G_{n_2,2}$, yields
\begin{eqnarray}
\label{sol_psi_l1}
\psi
=
\chi
-
3\sqrt{\frac{r_g}{r^3}}
C_{k_1}(t)
+
\frac{3r-2r_g}{r^2\sqrt{r_g}}
C_{k_r}(t)
+
r^2
\left(
\delta h-\delta k
\right)
+
\frac{Q_1(t)}{\sqrt{r}}
+
Q_2(t),
\end{eqnarray}
where $Q_1(t)$ and $Q_2(t)$ are free functions of time.
The compatibility of Eq.~\eqref{psih} and Eq.~\eqref{chik} then yields
\begin{eqnarray}
\label{sol_dh_l1}
\delta h
&=&
\delta k
+
\frac{1}{6r^4}
\Big[
2(5r-r_g)\sqrt{\frac{r}{r_g}}
C_{k_1}(t)
-
\sqrt{r_g}
C_{k_r}(t)
-
2r^{\frac{3}{2}}
Q_1(t)
-
r^2
Q_2(t)
+
6r^3
Q_3(t)
\nonumber
\\
&&
+
6 r^{\frac{7}{2}}
Q_4(t)
+
2r^2
C_{k_1}'(t)
+
2r^{\frac{3}{2}}
C_{k_r}'(t)
\Big],
\end{eqnarray}
where $Q_3(t)$ and $Q_4(t)$ are free functions of time.
After imposing Eqs.~\eqref{sol_n2_l1}, \eqref{sol_psi_l1}, and \eqref{sol_dh_l1}, the equations of motion for $\psi$, $\delta h$, $\chi$, and $\delta k$ coincide,
and hence we may focus on the equation of motion for $\psi$.

A combination of the equations of motion for $n_2$ and $m_2$ relates $\lambda_3$ to $\lambda_1$ and $\lambda_2$ as
\begin{eqnarray}
\label{sol_lambda3_ell1}
\lambda_3
&=&
\frac{2(1+C_0^2)\sqrt{r_g}}{C_0^2(1+C_0)\sqrt{r^7}}
\left(
\lambda_1
-r
\lambda_1'
-\frac{r^2}{2}
\lambda_1''
\right)
+
\frac{1}{r}
\lambda_2
\nonumber
\\
&-&
\frac{C_0 \alpha^2\left[
r
\left(
Q_2(t)
+
6r_g
\left(
Q_3(t)+\sqrt{r}Q_4(t)
\right)
-
12C_{k_1}'(t)
+
2\sqrt{r_g} Q_1'(t)
\right)
+
2\sqrt{r^3r_g}
Q_2'(t)
\right]}{2(1+C_0)(C_0c_1+c_2)(1+C_0^2\alpha^2)m^2\sqrt{r^7r_g}}.
\end{eqnarray}
Substituting Eq.~\eqref{sol_lambda3_ell1} back to the evolution equation for $n_2$, we obtain 
\begin{eqnarray}
\label{dot_chi}
\dot{\chi}
&=&
r^2
\dot{\delta k}
+
\sqrt{\frac{r^3}{r_g}}
\left(
r-r_g
\right)
\delta k'
+
\frac{5r-9r_g}
       {2}
\sqrt{
\frac{r}{r_g}
}
\delta k
-
\frac{r+r_g}{2\sqrt{r^3r_g}}
\chi
-
\frac{-r+r_g}
       {\sqrt{rr_g}}
\chi'
\nonumber\\
&&
+
\frac{-8r+7r_g}{r^3}
C_{k_1}(t)
+
\frac{2\left[r_g+C_0^2\alpha^2 (-3r+r_g)\right]}
       {\sqrt{r^3r_g}(1+C_0^2\alpha^2)}
C_{k_1}'(t)
+
\frac{3r^2-15rr_g+11r_g^2}
        {2r_g r^{\frac{7}{2}}}
C_{k_r}(t)
+
\frac{-3r+r_g}{r^2\sqrt{r_g}}
C_{k_r} '(t)
\nonumber\\
&&
+
\left(
\frac{11}{2}-\frac{6r}{r_g}
\right)h_1
-3
\left(1-\frac{r_g}{r}\right)h_r
+
\left(
4r
-
\frac{3r^2}{r_g}
-
r_g
\right)
 h_1'
+
\left(
-
\frac{3r^{\frac{3}{2}}}{\sqrt{r_g}}
+
\sqrt{rr_g} 
\right)
\dot{h}_1,
\nonumber
\\
&&
-
\frac{Q_2(t)
+
2
\left(
3r_gQ_3(t)
+
3\sqrt{r}r_gQ_4(t)
+
\sqrt{r_g}
Q_1'(t)
+
\sqrt{rr_g}
Q_2'(t)
\right)}{2\sqrt{rr_g}( 1+C_0^2\alpha^2)}.
\end{eqnarray}
Substituting Eq.~\eqref{dot_chi} and its derivatives with respect to $r$ into the equations of motion for $h_1$, $k_1$, $h_r$, and $k_r$,
we find that the equations of motion for $h_1$ and $k_1$ coincide, and similarly the equations of motion for $h_r$ and $k_r$ coincide, respectively.
This is because either $h_1$ or $k_1$ and either $h_r$ or $k_r$ can be eliminated by the gauge DOFs for the $\ell=1$ mode.
The consistency of the equations of motion for $h_1$ and $h_r$ then requires 
\begin{eqnarray}
\label{sol_lam1_l1}
\lambda_1(t)
&=&
\frac{Q_5(t)}{r^2}
+
r Q_6(t)
+
r^3 Q_7(t)
-
\frac{4C_0^3\alpha^2r^{\frac{3}{2}}\sqrt{r_g} 
\left(
3\sqrt{r_g} Q_4(t)
+
Q_2'(t)
\right)
}{21(1+C_0^2)m^2r_g(1+C_0^2\alpha^2)(C_0c_1+c_2) },
\end{eqnarray}
where $Q_5(t)$, $Q_6(t)$, and $Q_7(t)$ are free functions of time.
We note that $Q_5(t)$ and $Q_6(t)$ describe the solutions of the Laplace equation for $\ell=1$,
and without loss of generality we can impose that $Q_5(t)=0$ and $Q_6(t)=0$.

At this stage, the equations of motion $h_r$, $k_r$, $h_1$, and $k_1$ coincide.
Without loss of generality, we focus on  the equation of motion for $h_r$ which yields
\begin{eqnarray}
\label{sol_lam1_l2}
\lambda_2
&=&
Q_8(t)
-
\frac{10(1+C_0^2)\sqrt{rr_g}}
       {C_0^2 (1+C_0)}
Q_7(t)
\nonumber\\
&+&
\frac{C_0\alpha^2\left(
Q_2(t)
+
6r_g
Q_3(t)
+
6\sqrt{r}r_g
Q_4(t)
+
2\sqrt{r_g}
Q_1'(t)
+
2\sqrt{r r_g}
Q_2'(t)
-
12
C_{k_1}'(t)
\right)}
        {6(1+C_0) (C_0 c_1+c_2) (1+C_0^2\alpha^2)m^2 r^{\frac{3}{2}}\sqrt{r_g}}.
\end{eqnarray}
Requiring that $\lambda_1\to 0$ in the limit of $r\to \infty$, from Eq.~\eqref{sol_lam1_l1}, we impose that  $Q_7(t)=0$ and $Q_4(t)=-\frac{1}{3\sqrt{r_g}}Q_2'(t)$.
Then, we obtain $\lambda_1=0$.
Requiring that $\lambda_2\to 0$ in the limit of $r\to \infty$, from Eq.~\eqref{sol_lam1_l2}, we impose that $Q_8(t)=0$, and then obtain
\begin{eqnarray}
\lambda_2
&=&
\frac{C_0\alpha^2\left(
Q_2(t)
+
6r_g Q_3(t)
-
12
C_{k_1}'(t)
+
2\sqrt{r_g}
Q_1'(t)
\right)}
        {6(1+C_0) (C_0 c_1+c_2) (1+C_0^2\alpha^2)m^2 r^{\frac{3}{2}}\sqrt{r_g}},
\qquad 
\lambda_3
=
-\frac{2}{r}
\lambda_2.
\end{eqnarray}
By introducing the new variable $\phi$ to eliminate $\delta k$ by 
\begin{eqnarray}
\label{sol_dk_l1}
\delta k
=
\frac{\chi}{r^2}
+
\phi,
\end{eqnarray}
and substituting Eq.~\eqref{sol_dk_l1} into Eq.~\eqref{dot_chi}, $\chi$ is related to $\phi$ as
\begin{eqnarray}
\chi
&=&
\frac{r^{\frac{7}{2}}} {3\sqrt{r_g}}\dot{\phi}
+
\frac{r^3 (r-r_g)}{3r_g}\phi'
+
\frac{r^2(5r-9r_g)}{6r_g}\phi
+
\frac{-8r+7r_g}{3\sqrt{r^3r_g}} C_{k_1}(t)
+
\frac{3r^2-15rr_g+11r_g^2} {6r^2r_g^{\frac{3}{2}}}C_{k_r}(t)
+
\frac{-3r+r_g}{3\sqrt{r}r_g} C_{k_r}'(t)
\nonumber\\
&-&
\frac{rQ_2(t)+6rr_gQ_3(t)-4r_g C_{k_1}'(t)+12C_0^2r\alpha^2C_{k_1}'(t)-4C_0^2r_g\alpha^2C_{k_1}'(t)+2r\sqrt{r_g}Q_1'(t)}
        {6r_g(1+C_0^2\alpha^2)}
\nonumber\\
&+&
\frac{1}{6r^{\frac{11}{2}}r_g^{\frac{3}{2}}}
\Big(
r^7 (-12r+11r_g)h_1
+
6r^6r_g (-r+r_g)h_r
-
6r^9h_1'
+
8r^8 r_g h_1'
-
2r^7 r_g^2 h_1'
+
2r^{\frac{15}{2}}r_g^{\frac{3}{2}}
\dot{h}_1
-
6\sqrt{r^{17}r_g}
\dot{h}_1
\Big).
\end{eqnarray}
Since we have already employed all the equations of motion, in general $\phi$ is undetermined.
However, since $n_2$ and $m_2$ are given by 
\begin{eqnarray}
n_2
&=&
-
\frac{3C_{k_1}(t)}{r^2}
+
\frac{(r-4r_g)C_{k_r}(t)}
        {r^{\frac{5}{2}}r_g}
-
\frac{r^{\frac{3}{2}}}{3\sqrt{r_g}}
\left(
10\phi+11r\phi'+2r^2\phi''
\right)
\nonumber\\
&+&
\frac{1}{6r_g}
\Big(
\left(
36r-5r_g
\right)h_1
+18 r_g h_r
+
2r
\left(
(27r-5r_g)h_1'
+
6r_g h_r'
+
2r(3r-r_g)h_1''
\right)
\Big),
\\
\label{m2_ell1}
m_2&=&
-
\frac{4C_{k_1}(t)}{r^2}
+
\frac{\left(
2r-9r_g
\right)C_{k_r}(t)}
        {2r^{\frac{5}{2}}r_g}
-
\frac{
r^{\frac{3}{2}}}
{3\sqrt{r_g}}
\left(
10\phi+11r\phi'+2r^2\phi''
\right)
\nonumber\\
&+&
\frac{1}{6r_g}
\Big(
\left(
36r-5r_g
\right)h_1
+18 r_g h_r
+
2r
\left(
(27r-5r_g)h_1'
+
6r_g h_r'
+
2r(3r-r_g)h_1''
\right)
\Big).
\end{eqnarray}
we find that the gauge invariant (see Eqs. \eqref{rule_g_ell1} and \eqref{rule_f_ell1}) combination $n_2-m_2$ does not depend on $\phi$, $h_r$, and $h_1$. Because of the presence of the gauge degrees of freedom, $\Xi_r$ and $\Xi_1$ we may choose the gauge $h_1=0$ and $h_r=0$ as Eq.~\eqref{gauge_l1}.
Furthermore, as a particular solution we may choose 
\begin{eqnarray}
\label{phi_ell1}
\phi
=
0,
\end{eqnarray}
which avoids the growing terms in $n_2$ and $m_2$ as $r\to \infty$.

Below, we focus on the leading behavior of the perturbed metric components in the large distance limit $r\to \infty$ and impose their regularity.
The imposition of a sufficient number of boundary conditions in the limit of $r\to \infty$ will fix the remaining free functions of time, $Q_1(t)$, $Q_2(t)$, $Q_3(t)$, $C_{k_r}(t)$, and $C_{k_1}(t)$.
We will clarify the boundary conditions which are necessary to eliminate these free functions of time.

The leading terms in the  $(t,t)$-components of the metric perturbations which are obtained from linear combinations of $n_0$, $\frac{r_g}{r}n_1$, $\frac{r_g}{r}n_2$, $m_0$, $\frac{r_g}{r}m_1$, and $\frac{r_g}{r}m_2$ are proportional to $Q'_{2}(t) r$. 
In order to satisfy the regularity of both the sectors in the large distance $r\to\infty$, we require that the ${\cal O} (r)$ terms in the $(t,t)$-components of the perturbed metrics vanish, and hence impose
\begin{eqnarray}
\label{Q2Q10}
Q_2(t)=q_2.
\end{eqnarray} 
where $q_2$ is an integration constant.

The leading terms in the  $(t,r)$-components of the metric perturbations which are obtained from the linear combinations of $\sqrt{\frac{r_g}{r}}n_1$, $\sqrt{\frac{r_g}{r}}n_2$, $\sqrt{\frac{r_g}{r}}m_1$, and $\sqrt{\frac{r_g}{r}}m_2$ are given by the ${\cal O}(r^0)$ terms, which lead to the divergence of the ADM masses in both the sectors in the large distance limit $r\to \infty$.
The regularity of the ADM masses requires
\begin{eqnarray}
\label{Q3Q9}
&&
Q_3(t)=0,
\qquad
Q_1(t)
=
q_1
-
\frac{1}
       {2\sqrt{r_g}}
\left(
q_2t
+
12C_0^2\alpha^2C_{k_1}(t)
\right),
\end{eqnarray}
where $q_1$ is an integration constant.
The leading terms in the  $(r,r)$-components of the metric perturbations which are obtained from $n_2$ and $m_2$ are given by the  ${\cal O} \left(\frac{1}{r^{\frac{3}{2}}}\right)$, where the coefficients are proportional to $C_{k_r}(t)$.

The leading and sub-leading terms in the  $(t,\theta)$-components of the metric perturbations which are obtained from the linear combinations of $r^2h_t$, $r^2\sqrt{\frac{r_g}{r}}h_r$, $r^2k_t$, and $r^2\sqrt{\frac{r_g}{r}}k_r$ are given by the ${\cal O}(r^{\frac{1}{2}})$ and ${\cal O}(r^0)$ terms, respectively. 
For the asymptotic flatness of the spacetimes in both the sectors, to eliminate ${\cal O}(r^{\frac{1}{2}})$ term requires
\begin{eqnarray}
\label{cond_ckr}
C_{k_r}(t)=c_{k_r},
\end{eqnarray}
where $c_{k_r}$ is an integration constant.
Similarly, eliminating the ${\cal O}(r^0)$ terms in the  $(t,\theta)$-components of the metric perturbations requires that
\begin{eqnarray}
\label{cond_ck1}
C_{k_1}(t)
=
\frac{q_2}{2}
t
+c_{k_1},
\qquad 
c_{k_r}
=
-\frac{2}{3}
q_2r_g^{\frac{3}{2}},
\end{eqnarray}
where $c_{k_1}$ is an integration constant.
The $(r,\theta)$-components of the metric perturbations obtained from $r h_r$ and $r k_r$ automatically vanish and need not to be considered in the rest.
The leading terms in the angular components of the perturbed metric are proportional to ${\cal O} (r^0)$
which are suppressed by the factor $\frac{1}{r^2}$ compared to the background metrics.

After imposing Eq.~\eqref{cond_ckr}, the ${\cal O} (r^0)$ terms of the  $(t,t)$-components of the metric perturbations automatically vanish.
Similarly, after imposing Eq.~\eqref{cond_ckr}, the ${\cal O} \left(\frac{1}{r^{\frac{1}{2}}}\right)$ terms of the  $(t,r)$-components of the metric perturbations automatically vanish.
Then, the ${\cal O} \left(\frac{1}{r^{\frac{1}{2}}}\right)$ terms in the $(t,t)$-components of the metric perturbations
are proportional to $\frac{q_2}{\sqrt{r}}$.
Requiring that these terms vanish as well imposes
\begin{eqnarray}
\label{q2q10}
q_2=0,
\end{eqnarray}
and from Eq. \eqref{cond_ck1}  we obtain that $c_{k_r}=0$.
The ${\cal O} \left(\frac{1}{r}\right)$ terms in the $(t,r)$-components of the metric perturbations vanish.
With Eq.~\eqref{q2q10}, the ${\cal O} \left(\frac{1}{r}\right)$ and ${\cal O} \left(\frac{1}{r^{\frac{3}{2}}}\right)$ terms in the $(t,t)$-components of the metric perturbations automatically vanish.

The next-order terms in the  $(t,\theta)$-components of the metric perturbations  are given by the ${\cal O}\left(\frac{1}{r^{\frac{1}{2}}}\right)$ terms.
Requiring that these terms vanish imposes
\begin{eqnarray}
\label{cond_Q1q9}
q_1=0,
\qquad 
c_{k_1}=0.
\end{eqnarray}
After imposing Eq.~\eqref{cond_Q1q9}, the ${\cal O} \left(\frac{1}{r^{3/2}}\right)$ terms in the $(t,r)$-components of the metric perturbations automatically vanish.
As a consequence, all the components of the $\ell=1$ perturbations vanish.

The manipulations to reduce the equations of motion for the $\ell=1$ mode are summarized in Table~\ref{tablel1}.

\begin{table}[htbp]
\begin{center}
\begin{tabular}{ |c|c|c|c| } 
\hline
 Manipulation & Output & Remark & \# of  independent variables \\
\hline
\hline
Derive EOMs         &                     & & 18\\ 
Symmetry            & $h_2=0$ and $k_2=0$ & & 16\\
Gauge-fixing        & $h_1=0$ and $h_r=0$ & & 14\\
EOM for $n_0$       & Eliminate $n_0$     & & 13\\
EOM for $m_0$       & Eliminate $m_0$     & & 12\\ 
EOM for $\lambda_1$ & Eliminate $m_2$     & & 11\\
EOM for $\lambda_2$ and $\lambda_3$ & Eliminate $k_r$ and $k_1$ & $C_{k_1}(t)$, $C_{k_r}(t)$ & 9\\
Combine EOM for $\psi$ and EOM for $\delta h$ & Eliminate $n_2$ & & 8\\
Combine EOM for $\chi$ and EOM for $\delta k$ & Eliminate $\psi$ & $Q_1(t)$, $Q_2(t)$ & 7\\
Combine EOM for $\psi$ and EOM for $\chi$  & Eliminate $\delta h$ & $Q_3(t)$, $Q_4(t)$ & 6\\
Combine EOM for $n_2$ and EOM for $m_2$ & Eliminate $\lambda_3$ & & 5\\
Combine EOM for $h_1$ and EOM for $h_r$ & Eliminate $\lambda_1$ & $Q_7(t)$ & 4\\
EOM for $h_r$      & Eliminate $\lambda_2$ & $Q_8$(t) &3 \\
Regularity of $\lambda_1$ at $r\rightarrow\infty$ & Set $\lambda_1 = 0$ & $Q_7(t) = 0$, fix $Q_4(t)$ & 3\\
Regularity of $\lambda_2$ at $r\rightarrow\infty$ & & $Q_8(t) = 0$ & 3\\
EOM for  $n_2$ &
&  & 2 
\\
Introduce $\phi$ (Eq.~\eqref{sol_dk_l1})
 &
Eliminate $\chi$
& 
& 1 
\\
Gauge invariance of $n_2-m_2$
 &
$\phi=0$ & & 1 \\
Set $\lambda_0$ &$\lambda_0 = 0$  & & 0 
\\
Eliminating ${\cal O} (r)$ in the $(t,t)$ components
&
Eq. \eqref{Q2Q10}
&
fix $Q_2(t)$ 
&
0
\\
Eliminating ${\cal O} (r^0)$ in the $(t,r)$ components
&
Eq. \eqref{Q3Q9}
&
fix $Q_3(t)$ and $Q_1(t)$
&
0
\\
Eliminating ${\cal O} (\sqrt{r})$ and ${\cal O} (r^0)$ in the $(t,\theta)$ components
&
Eq. \eqref{cond_ckr} and \eqref{cond_ck1}
&
fix $C_{k_1}(t)$ and $C_{k_1}(t)$
&
0
\\
Eliminating ${\cal O} (\frac{1}{\sqrt{r}})$ in the $(t,t)$ components
&
Eq. \eqref{q2q10}
&
fix $q_2$
&
0
\\
Eliminating ${\cal O} (\frac{1}{\sqrt{r}})$ in the $(t,\theta)$ components
&
Eq. \eqref{cond_Q1q9}
&
fix $q_1$ and $c_{k_1}$
&
0
\\
\hline
\end{tabular}
\caption{
The manipulations to reduce the equations of motion for the $\ell=1$ mode.
}
\label{tablel1}
\end{center}
\end{table}

\section{The solutions of the higher multipolar perturbations}
\label{sec6}

In this section, we focus on the $\ell\geq 2$ modes.

\subsection{The case of the two copies of GR}

Again, as a point of comparison, we first review the case of  the two copies of GR, i.e., the case of $m=0$.
For the modes $\ell \geq 2$, after deriving the equations of motion for $m=0$, under the two copies of the four-dimensional diffeomorphism invariance, for the perturbed metrics \eqref{metric_g_ell} and \eqref{metric_f_ell} we fix the gauge as 
\begin{eqnarray}
{h}_t=-\sqrt{\frac{r_g}{r^3}}{h}_r,
\qquad
h_1=0,
\qquad 
h_2=0,
\qquad 
k_t=-
\sqrt{\frac{r_f}{r^3}}{k}_r,
\qquad
k_1=0,
\qquad 
k_2=0.
\end{eqnarray}
We then introduce the master variables $\psi$ and $\chi$ as
\begin{eqnarray}
\label{rel_n1}
n_1
&=&
\frac{1}{2r_g}
\Big[
\ell (\ell+1)
\left(
r-r_g
\right)
h_r
+
2r_g n_0
-
r n_2
+
\kappa^2
\sqrt{\ell (\ell+1)}
\frac{(\ell^2+\ell-2)r+3r_g}
       {r}
\psi
\Big],
\nonumber\\
\label{rel_m1}
m_1
&=&
\frac{1}{2r_f}
\Big[
\ell (\ell+1)
\left(
b^2r-r_f
\right)
k_r
+
2r_f m_0
-
b^2r m_2
+
\frac{\kappa^2}{\alpha^2}
\sqrt{\ell (\ell+1)}
\frac{(\ell^2+\ell-2)b^2r+3r_f}
       {r}
\chi
\Big].
\end{eqnarray}
The equations of motion for $n_0$ and $m_0$ relate $n_0$ and $m_0$ to other variables, respectively.
The combination of the equations of motion for $h_1$ and $h_2$ is used to eliminate $n_2$.
Similarly, the combination of the equations of motion for $k_1$ and $k_2$ is used to eliminate $m_2$.
Then, the combination of the equations of motion for $\psi$ and $h_1$ yields
\begin{eqnarray}
&&
\label{master_gr_h}
\ddot{\psi}
-2\sqrt{\frac{r_g}{r}}\dot{\psi}'
-\left(1-\frac{r_g}{r}\right)\psi''
+\frac{1}{2}\sqrt{\frac{r_g}{r^3}}\dot{\psi}
-\frac{r_g}{r^2}\psi'
\nonumber\\
&&
+
\frac{\ell(\ell+1) (\ell^2+\ell-2)^2 r^3+3 (\ell^2+\ell-2)^2r^2r_g+9(\ell^2+\ell-2)rr_g^2+9r_g^3}
        {r^3\left((\ell^2+\ell-2)r+3r_g\right)^2}
\psi
=0.
\end{eqnarray}
Similarly, the combination of the equations of motion for $\chi$ and $k_1$ yields
\begin{eqnarray}
\label{master_gr_k}
&&
\ddot{\chi}
-2\sqrt{\frac{r_f}{r}}\dot{\chi}'
-\left(b^2-\frac{r_f}{r}\right)\chi''
+\frac{1}{2}\sqrt{\frac{r_f}{r^3}}\dot{\chi}
-\frac{r_f}{r^2}\chi'
\nonumber\\
&&
-
\frac{\ell(\ell+1) (\ell^2+\ell-2)^2b^6 r^3+3b^4 (\ell^2+\ell-2)^2r^2r_f+9b^2(\ell^2+\ell-2)rr_f^2+9r_f^3}
        {r^3\left(b^2(\ell^2+\ell-2)r+3r_f\right)^2}
\chi
=0.
\end{eqnarray}
With Eqs.~\eqref{master_gr_h} or \eqref{master_gr_k}, we confirm that the rest of the equations of motion in the physical and fiducial sectors are satisfied, respectively.
Thus, $\psi$ and $\chi$ play the role of the master variables in the physical and fiducial sectors, respectively.

\subsection{The case of MTBG}

We then focus on the $\ell\geq 2$ modes in the self-accelerating branch of MTBG.
In order to eliminate the dependence on $h_t$ and $k_t$, we introduce the new variables $\delta h$ and $\delta k$ as Eqs.~\eqref{deltahk}.
We also introduce the new variables $\psi$ and $\chi$ to replace $n_1$ and $m_1$ by
\begin{eqnarray}
\label{rel_n1}
n_1
&=&
\frac{1}{8r_g}
\Big[
2\ell (\ell+1)
\sqrt{r^3 r_g} \delta h
+
2
\left(
6 r-5r_g
\right)
h_1
-
\ell
(\ell+1)
\left(
6r-5r_g
\right)
h_2
+
4\ell (\ell+1)
\left(
r-r_g
\right)
h_r
\nonumber\\
&&
+
4r (r-r_g)
h_1'
-
2\ell (\ell+1)r (r-r_g)
h_2'
+
4\sqrt{r^3 r_g}
\dot{h}_1
-
2\ell(\ell+1)
\sqrt{r^3 r_g}
\dot{h}_2
\nonumber\\
&&
+
8r_g n_0
-
4r n_2
+
\frac{4\kappa^2}{r}
\sqrt{\ell(\ell+1)}
\left(
(\ell^2+\ell-2)r+3r_g
\right)
\psi
\Big],
\nonumber\\
\label{rel_m1}
m_1
&=&
\frac{1}{8r_f}
\Big[
2\ell (\ell+1)
\sqrt{r^3 r_f} \delta k
+
2
\left(
6 b^2 r-5r_f
\right)
k_1
-
\ell
(\ell+1)
\left(
6b^2r-5r_f
\right)
k_2
+
4\ell (\ell+1)
\left(
b^2r-r_f
\right)
k_r
\nonumber\\
&&
+
4r (b^2r-r_f)
k_1'
-
2\ell (\ell+1)r (b^2r-r_f)
k_2'
+
4\sqrt{r^3 r_f}
\dot{k}_1
-
2\ell(\ell+1)
\sqrt{r^3 r_f}
\dot{k}_2
\nonumber\\
&&
+
8r_f m_0
-
4b^2r m_2
+
\frac{4\kappa^2}{\alpha^2 r}
\sqrt{\ell(\ell+1)}
\left(
(\ell^2+\ell-2)b^2r+3r_f
\right)
\chi
\Big].
\end{eqnarray}
After deriving the totally eighteen components of the equations of motion for the eighteen variables, we fix $h_1$ and $h_2$ to $0$ by the gauge conditions \eqref{h1h20}.
The equation of motion for $\lambda_0$ fixes $m_2$ as
\begin{eqnarray}
\label{rel_m2}
m_2
&=&
-\frac{1}{2}k_1
+
\frac{\ell (\ell+1)}{4}
k_2
+
n_2.
\end{eqnarray}
The equation of  motion for $\lambda_1$ is not independent of that for $\lambda_0$.
The equation of motion for $\lambda_2$ fixes $k_r$ as
\begin{eqnarray}
k_r
=
h_r
+
\frac{1}{2\ell (\ell+1)}
\left[
-6k_1
-4rk_1'
+
\ell(\ell+1)
(3k_2+2rk_2')
\right].
\end{eqnarray}
The equations of motion for $n_0$ and $m_0$ fix $n_0$ and $m_0$, respectively.
The compatibility of the equations of motion for $h_1$ and $h_2$ fixes $n_2$.
Similarly, 
the compatibility of the equations of motion for $k_1$ and $k_2$ fixes $\lambda_3$.
The equation of motion for $\lambda_3$ provides the constraint relation
\begin{eqnarray}
\label{con_k2}
k_2''
=
\frac{2(\ell^2+\ell+18)k_1+44rk_1'+\ell (\ell+1)\left[(\ell^2+\ell-22)k_2-22rk_2'\right]+8r^2k_1''}
       {4\ell (\ell+1)r^2}.
\end{eqnarray}
The combination of the equations of motion for $h_r$ and $k_r$ provides the evolution equation for $k_2$ 
\begin{eqnarray}
\label{eqs_k2}
\ddot{k}_2=G_{k_2},
\end{eqnarray} 
where $G_{k_2}$ represents at most first-order time derivative terms of the perturbation variables.
The compatibility of the equations of motion for $\delta h$ and $\delta k$ provides the constraint relation
\begin{eqnarray}
\label{con_dk}
\delta k''=H_{\delta k},
\end{eqnarray}
where $H_{\delta k}$ includes at most first-order time derivative terms of the perturbation variables.
The compatibility of the equations of motion for $\psi$ and $\chi$ provides  the constraint relation
\begin{eqnarray}
\label{con_chi}
\chi''=H_{\chi},
\end{eqnarray}
where $H_{\chi}$ includes at most first-order time and radial derivative terms of the perturbation variables.
The equation of motion for $n_2$ fixes $\lambda_2$.
The compatibility of the equations of motion for $m_2$ and $n_2$ provides a constraint relation denoted by 
\begin{eqnarray}
{\cal C}_1=0,
\label{c1}
\end{eqnarray}
which includes at most first-order time derivative terms of the perturbation variables.
The compatibility of the equations of motion for $h_1$ and $k_1$ provides
\begin{eqnarray}
\label{con_k1}
k_1''=H_{k_1},
\end{eqnarray} 
where $H_{k_1}$ contains at most first-order time derivative terms including $\dot{\psi}''$, $\dot{\chi}''$, $\dot{\delta h}''$, $\dot{\delta k}''$,  ${\delta h}''$, ${\delta k}''$, $\psi''$, $h_r''$, $\lambda_1'''$, and $\lambda_1''$.

With use of the constraint \eqref{c1}, the equations of motion for $h_1$, $\delta h$, and $\psi$ lead to the evolution equations 
\begin{eqnarray}
{\cal E}_1=0, 
\qquad 
{\cal E}_2=0,
\qquad 
 {\cal E}_3=0, 
\label{e1e2e3}
\end{eqnarray}
which include at most second-order time derivative terms.
We find that the evolution equations in Eq.~\eqref{e1e2e3} are degenerate with respect to the second-order derivatives of $t$ and can be rewritten as the evolution equation of 
\begin{eqnarray}
\label{dynamical}
&&
\Upsilon:=
2
\kappa^2
\sqrt{\frac{r}{r_g}}
\left(
(\ell^2+\ell-2)r
+
3r_g
\right)
\left(
\psi
+
C_0^2
\chi
\right)
\nonumber\\
&&-
\sqrt{\ell (\ell+1)}
r^4
\left(
\delta h
+
C_0^2\alpha^2 
\delta k
\right)
+
2
\sqrt{\frac{1}{\ell(\ell+1)}}
C_0^2\alpha^2 
r^2
\sqrt{\frac{r}{r_g}}
\left(
\frac{\ell^2+\ell+4}{2}
r-r_g
\right)
k_1.
\end{eqnarray}
The equation of motion for $h_r$ leads to the constraint relation
\begin{eqnarray}
{\cal C}_2=0,
\label{c2}
\end{eqnarray}
which includes at most first-order time derivative terms of the perturbation variables.
In the case of two copies of GR with $m=0$, $\psi$ and $\chi$ play the role of the master variables in each sector.
Instead, in the case of MTBG with $m\neq 0$, the evolution of the perturbations in the two sectors are coupled to each other.

The degeneracy between ${\cal E}_1=0$ and ${\cal E}_2=0$ in Eqs.~\eqref{e1e2e3} leads to the constraint relation
\begin{eqnarray}
\label{con_psi}
\psi''=H_{\psi},
\end{eqnarray}
where $H_{\psi}$ represents at most the first-order time derivative terms of the perturbation variables.
Similarly,  the degeneracy between ${\cal E}_2=0$ and ${\cal E}_3=0$ in Eqs.~\eqref{e1e2e3} reduces to the constraint relation Eq.~\eqref{c1} and does not produce any more constraint.
The manipulations to reduce the equations of motion for the $\ell\geq 2$ modes are summarized in Table~\ref{tablel2}.
In the effectively massless case with $b=1$ and $r_f=r_g$, $\lambda_0$ does not appear and hence without loss of generality we may set 
\footnote{
$\lambda_0$ may appear in higher-order perturbations even in the effectively massless case.
In such a case,  imposing the regularity of higher-order perturbations may fix $\lambda_0$ uniquely, although this is beyond the scope of the present paper.}
\begin{eqnarray}
\lambda_0=0.
\end{eqnarray}
\begin{table}[htbp]
\begin{center}
\begin{tabular}{ |c|c|c|c| } 
\hline
 Manipulation & Output & Remark & \# of  independent variables \\
\hline
\hline
Derive EOMs &  & & 18\\ 
Fix the gauge  & $h_1=0$ and $h_2=0$ & & 16\\ 
EOM for $\lambda_0$  & Eliminate $m_2$&  EOM for $\lambda_1$ is not independent & 15\\ 
EOM for $\lambda_2$  & Eliminate $k_r$ &     &14 \\
EOM for $n_0$   & Eliminate $n_0$ & &13 \\
EOM for $m_0$  & Eliminate $m_0$& &12 \\
Combine EOM for $h_1$ and EOM for $h_2$  & Eliminate $n_2$ & & 11\\
Combine EOM for $k_1$ and EOM for $k_2$  & Eliminate $\lambda_3$ &  & 10 \\
EOM for $\lambda_3$  & Eq. \eqref{con_k2} &1st constraint &9 \\
Combine EOM for $h_r$ and EOM for $k_r$ & Eq. \eqref{eqs_k2} & 1st evolution equation &8 \\
Combine EOM for $\delta h$ and EOM for  $\delta k$ & Eq. \eqref{con_dk} & 2nd constraint &7 \\
Combine EOM for $\psi$ and EOM for  $\chi$ & Eq. \eqref{con_chi} & 3rd constraint & 6\\
Combine EOM for $n_2$ and EOM for $m_2$ & ${\cal C}_1=0$ & 4th constraint &5\\
EOM for  $n_2$ (or EOM for $m_2$)  & Eliminate $\lambda_2$ &  & 4 \\
Combine EOM for $h_1$ and EOM for $k_1$  &Eq.~\eqref{con_k1} & 5th constraint & 3 \\
EOM for $h_1$ (or EOM for $k_1$)  & ${\cal E}_1=0$ & 2nd evolution equation & 2 \\
EOM for $h_r$ (or EOM for $k_r$)  & ${\cal C}_2=0$ & 6th constraint & 1\\
EOM for $\delta h$ (or EOM for $\delta k$)  & ${\cal E}_2=0$ & Degenerate to ${\cal E}_1=0$  & 1  \\
EOM for $\psi$ (or EOM for $\chi$)  & ${\cal E}_3=0$ &   Degenerate to ${\cal E}_1=0$ and ${\cal E}_2=0$& 1   \\
Take the difference between ${\cal E}_1=0$ and ${\cal E}_2=0$  & Eq. \eqref{con_psi} & 7th constraint &  0 \\
Take the difference between ${\cal E}_2=0$ and ${\cal E}_3=0$  & ${\cal C}_1=0$ &  Not an independent constraint &  0 \\
\hline
\end{tabular}
\caption{
The manipulations to reduce the equations of motion for the $\ell\geq 2$ modes.
}
\label{tablel2}
\end{center}
\end{table}
Each of the degenerate evolution equations ${\cal E}_1=0$ (${\cal E}_2=0$, or ${\cal E}_3=0$) still contains the terms with two time and one radial derivatives as $\ddot{\psi}'$.
However, by combining ${\cal E}_3=0$ with $\dot{{\cal C}}_1'=0$, all the two time and one radial derivative terms cancel and  the resultant evolution equation
\begin{eqnarray}
\tilde{\cal E}_3=0,
\label{te3}
\end{eqnarray}
is purely a second-order derivative equation with respect to time.

We then replace $\delta h$ with $\Upsilon$ using Eq.~\eqref{dynamical}.
We also relate $h_r$ to other variables through Eq.~\eqref{c1}.
Then, the equation \eqref{te3} reduces to 
\begin{eqnarray}
\bar{\cal E}_3=0,
\label{be3}
\end{eqnarray}
which has the structure
\begin{eqnarray}
\ddot{k}_2
+\left(-1+\frac{r_g}{r}\right)k_2''
+
\frac{2r_g^2}
        {C_0^2\alpha^2 (5r_g-6r)}
\left(
\frac{r}{r_g}
\right)^{\frac{3}{2}}
\left[
\ddot{\Upsilon}
+\left(-1+\frac{r_g}{r}\right)\Upsilon''
\right]
=
R_{\bar{\cal E}_3},
\label{evol_k2Psi2}
\end{eqnarray}
where the remaining terms $R_{\bar{\cal E}_3}$ do not contain second-order derivative terms with respect to time.
We note that the evolution equation \eqref{eqs_k2} is not independent of Eq.~\eqref{be3}.
After eliminating $h_r$ with Eq.~\eqref{c1}, the constraint Eq.~\eqref{c2} now turns into an evolution equation,
because the $\dot{h}_r$ term in the original \eqref{c2} turns into second-order derivative terms with respect to time.

After eliminating $h_r$ with Eq.~\eqref{c1}, 
the equations \eqref{con_dk}, \eqref{con_chi}, and \eqref{con_psi} finally reduce to the elliptic equations
\begin{eqnarray}
&&
\label{conspsi2}
\psi''=K_{\psi},
\\
&&
\label{conschi2}
\chi''=K_{\chi},
\\
&&
k_1''
=
-
\frac{2(\ell^2+\ell+18)k_1+44rk_1'+\ell (\ell+1)\left[(\ell^2+\ell-22)k_2-22rk_2'\right]-4\ell (\ell+1)r^2 k_2''}
       {8r^2},
\end{eqnarray}
where the last equation coincides with Eq.~\eqref{con_k2}, and $K_\psi$ and $K_\chi$ in  Eqs.~\eqref{conspsi2} and \eqref{conschi2} do not contain second-order derivative terms with respect to $r$ besides $k_2''$ and $\Upsilon''$.
The constraint relation \eqref{c2} now turns to the evolution equation
\begin{eqnarray}
\bar{\cal C}_2=0,
\label{bc2}
\end{eqnarray}
which has the structure
\begin{eqnarray}
\label{evol_k2Psi22}
\ddot{k}_2
+\left(-1+\frac{r_g}{r}\right)k_2''
+
\frac{r_g^2}
        {C_0^2\alpha^2 (r_g-\frac{\ell^2+\ell+4}{2}r)}
\left(
\frac{r}{r_g}
\right)^{\frac{3}{2}}
\left(
\ddot{\Upsilon}
+\left(-1+\frac{r_g}{r}\right)\Upsilon''
\right)
=
R_{\bar{\cal C}_2},
\end{eqnarray}
where the remaining terms $R_{\bar{\cal C}_2}$ do not contain second-order derivative terms with respect to time.
The combination of \eqref{evol_k2Psi2} and \eqref{evol_k2Psi22} leads to the individual evolution equations for $\Upsilon$ and $k_2$, which are schematically given by 
\begin{eqnarray}
\label{evol_upsilon}
\ddot{\Upsilon}
+\left(-1+\frac{r_g}{r}\right)\Upsilon''=R_{\Upsilon},
\\
\label{evol_k2}
\ddot{k}_2+\left(-1+\frac{r_g}{r}\right)k_2''=R_{k_2},
\end{eqnarray}
where $R_{k_2}$ and $R_{\Upsilon}$ are given by the linear combinations of $R_{\bar{\cal E}_3}$ and $R_{\bar{\cal C}_2}$.
These equations tell that in the case of the effectively massless case with $b=1$ and $r_f=r_g$ the two modes $\Upsilon$ and $k_2$ propagate with the speed of light in the radial directions.

The equation \eqref{con_k1} reduces to the constraint
\begin{eqnarray}
\bar{\cal C}_3=0,
\label{bc3}
\end{eqnarray}
which includes $\lambda_1 ''''$ and $\ddot{k}_2$.
After eliminating $\ddot{k}_2$ by Eq.~\eqref{evol_k2}, the equation \eqref{bc3} can be written solely as the fourth-order differential equation with respect to $r$ for $\lambda_1$
\begin{eqnarray}
\label{lam1_eq}
2r^4\lambda_1''''
+
5r^3 \lambda_1'''
-\frac{12+5\ell(\ell+1)}{2} r^2 \lambda_1''
+
6\left(\ell^2+\ell+1\right) r \lambda_1'
+
\frac{1}{2}
(\ell-4)
\ell 
(\ell+1)
(\ell+5)
\lambda_1
=0,
\end{eqnarray}
whose general solution is given by 
\begin{eqnarray}
\label{sol_lambda1}
\lambda_1
=
C_{\lambda_1,1}(t)
 r^{-\frac{\ell-4}{2}}
+
C_{\lambda_1,2}(t) 
r^{\frac{\ell+5}{2}}
+
C_{\lambda_1,3}(t)
 r^{-\ell-1}
+
C_{\lambda_1,4}(t)
 r^{\ell},
\end{eqnarray}
where $C_{\lambda_1,1(t)}$, $C_{\lambda_1,2}(t)$, $C_{\lambda_1,3}(t)$, and $C_{\lambda_1,4}(t)$ are functions of $t$.
Depending on the boundary conditions, these free functions should be chosen appropriately.
The fact that in the even-parity sector the equation for the instantaneous mode from the Lagrange multipliers is independent of the other modes is reminiscent of the case of the odd-parity perturbations in the effectively massless case  where the equation for the instantaneous mode $\Lambda$ associated with the Lagrange multiplier is independent of the other variables and can be solved analytically (See Eq.~(117) in Ref.~\cite{Minamitsuji:2023lvi}).
We note that the solutions of $C_{\lambda_1,3}(t)$ and $C_{\lambda_1,4}(t)$ are those of the Laplace equation in the three-dimensional Euclid space for the $\ell$ mode.
Plugging Eq.~\eqref{sol_lambda1} into the other independent equations,  $C_{\lambda_1,3}(t)$ and $C_{\lambda_1,4}(t)$ do not contribute to the remaining equation of motion.
Thus, $\lambda_1$ contains only one physical instantaneous mode, in spite of the fact that it follows the fourth-order differential equation \eqref{lam1_eq}.

In Eqs.~\eqref{con_k2}, \eqref{conspsi2}, and \eqref{conschi2}, the right-hand sides still contain $\Upsilon''$, $k_2''$, and $\delta k''$.
We introduce the new variables $\bar{\psi}$, $\bar{\chi}$, and $\bar{k}_1$ by 
\begin{eqnarray}
\psi&=&\bar{\psi}
+
\frac{\sqrt{2+3(\ell-1)+(\ell-1)^2}}
       {2\left(1+C_0^2\alpha^2\right)
\left[3r_g+(\ell^2+\ell-2)r\right]\kappa^2}
r^{\frac{7}{2}}
r_g^{\frac{1}{2}}
\left[\left(1+C_0^2\alpha^2\right) \delta k + \Upsilon\right]
\nonumber\\
&&
+
\frac{C_0^2\alpha^2}{8\left(1+C_0^2\alpha^2\right)}
\frac{\sqrt{2+3(\ell-1)+(\ell-1)^2}\left\{
4r_g-2\left[6+3(\ell-1)+(\ell-1)^2\right]r
\right\} }
       {\left[3r_g+(\ell^2+\ell-2)r\right]\kappa^2}
r^2
k_2,
\\
\chi&=&\bar{\chi}
+
\alpha^2
\frac{\sqrt{2+3(\ell-1)+(\ell-1)^2}}
       {2\left(1+C_0^2\alpha^2\right)\left[3r_g+(\ell^2+\ell-2)r\right]\kappa^2}
r^{\frac{7}{2}}
r_g^{\frac{1}{2}}
\left[\left(1+C_0^2\alpha^2\right)\delta k + \Upsilon\right]
\nonumber\\
&&
+
\frac{C_0^2\alpha^4}{8\left(1+C_0^2\alpha^2\right)}
\frac{\sqrt{2+3(\ell-1)+(\ell-1)^2}\left\{
4r_g-2\left[6+3(\ell-1)+(\ell-1)^2\right]r
\right\} }
       {\left[3r_g+(\ell^2+\ell-2)r\right]\kappa^2}
r^2
k_2,
\\
k_1&=&\bar{k}_1+\frac{\ell(\ell+1)}{2}k_2,
\end{eqnarray}
and eliminate $\psi$, $\chi$, and $k_1$, so that Eqs.~\eqref{con_k2}, \eqref{conspsi2}, and \eqref{conschi2} become purely the second-order spatial equations for $\bar{\psi}$, $\bar{\chi}$, and $\bar{k}_1$.
We note that after introducing $\bar{\psi}$, $\bar{\chi}$, and $\bar{k}_1$, the dependence on $\delta k$ never shows up in the equations of motion, and turns out to be a gauge mode.

By introducing $\bar{\psi}$, $\bar{\chi}$, and $\bar{k}_1$, the master equations for the two dynamical modes \eqref{evol_upsilon} and \eqref{evol_k2} turn to be
\begin{eqnarray}
\label{eq2_upsilon}
\ddot{\Upsilon}
-2\sqrt{\frac{r_g}{r}}\dot{\Upsilon}'
+\left(-1+\frac{r_g}{r}\right)\Upsilon''=\bar{R}_{\Upsilon}.
\\
\label{eq2_k2}
\ddot{k}_2-2\sqrt{\frac{r_g}{r}}\dot{k}_2'+\left(-1+\frac{r_g}{r}\right)k_2''=\bar{R}_{k_2},
\end{eqnarray}
where $\bar{R}_{\Upsilon}$ and $\bar{R}_{k_2}$ contain at most first-order derivatives with respect to the time and radial coordinates.
We note that the left-hand sides of \eqref{eq2_upsilon} and \eqref{eq2_k2} correspond to the GR operators in the even-parity perturbations.
Eqs.~\eqref{con_k2}, \eqref{conspsi2}, \eqref{conschi2},  and \eqref{sol_lambda1} now reduce to three elliplitic equations and the physical solution for $\lambda_1$,
\begin{eqnarray}
&&
{\bar k}_1''
=
-
\frac{2(\ell^2+\ell+18)\bar{k}_1+44r\bar{k}_1'+2(\ell+2) (\ell+1)\ell (\ell-1)k_2}
       {8r^2},
\\
&&
\bar{\psi}''=\bar{K}_{\psi},
\\
&&
\bar{\chi}''=\bar{K}_{\chi},
\\
&&
\lambda_1
=
C_{\lambda_1,1}(t)
 r^{-\frac{\ell-4}{2}}
+
C_{\lambda_1,2}(t) 
r^{\frac{\ell+5}{2}},
\end{eqnarray}
where $\bar{K}_{\psi}$ and $\bar{K}_{\chi}$ include the at most the first-order radial derivatives.
Thus, in the effectively massless case, there are two dynamical modes $\Upsilon$ and $k_2$, while there are four instantaneous modes from $\bar{k}_1$, $\bar{\psi}$, $\bar{\chi}$, and $\lambda_1$.
As in the case of the odd-parity sector~\cite{Minamitsuji:2023lvi}, in the effectively massless case the two modes in the even-parity sector propagate with the speeds of light at least in the radial direction.
Combined with the analysis of the odd-parity sector~\cite{Minamitsuji:2023lvi}, in the effectively massless case there are four propagating DOFs and six instantaneous DOFs for each of the $\ell \geq 2$ modes.

If the effectively massless condition is relaxed, there may be more instantaneous modes, while the number of propagating modes should remain the same because of  the structure of MTBG.
Also as in the case of the odd-parity sector~\cite{Minamitsuji:2023lvi}, we expect that if the condition is relaxed the speeds of the two propagating modes would differ from each other.
We also confirm that the squared angular propagation speeds of the two evolution modes $\Upsilon$ and $k_2$ read from Eqs.~\eqref{evol_upsilon} and \eqref{evol_k2} are positive, indicating that at least in the effectively massless case there is no instability in the angular directions in the large distance limits.

\section{Conclusions}
\label{sec7}

As a continuation of the previous works \cite{Minamitsuji:2022vfv,Minamitsuji:2023lvi}, we have studied even-parity perturbations about static and spherically symmetric Schwarzschild solutions in the self-accelerating branch of MTBG.
Before performing the analysis of even-parity perturbations, in Sec.~\ref{sec2} we have reviewed the Minkowski solutions in the self-accelerating branch of MTBG as the limit of the vanishing effective cosmological constants of the de~Sitter solutions written in the spatially-flat FLRW coordinates.
We have shown that in general the physical and fiducial sectors do not share the same Minkowski vacuum because under the joint foliation-preserving diffeomorphism invariance the time coordinate cannot describe the proper time in both the physical and fiducial sectors.
In order to share the same Minkowski vacuum in both the sectors, the free function which measures the difference in the proper times between the two sectors has to be constant.
In Sec.~\ref{sec3}, we have reviewed the even-parity perturbations on the Schwarzschild solutions in the spatially-flat coordinates in the self-accelerating branch of MTBG.
For each of the $\ell=0$, $1$, and $\geq 2$ modes, we have clarified the gauge transformations under the joint three-dimensional diffeomorphism transformation.

In Sec.~\ref{sec4}, we have investigated the solution for the $\ell=0$ mode.
As expected from the structure of MTBG, we confirmed that there is no propagating DOF for $\ell=0$.
The general solution to the $\ell=0$ mode allows that the mass of a BH in each sector varies with time, while the summation of masses in both the sectors remains constant.
However, by requiring that the total mass of matter in each sector inside a star is constant before gravitational collapse, the mass of the BH in each sector has to be constant, which fixes one of the free functions of time  to be a constant.
Moreover, the requirement that the asymptotic regions of the spacetimes in both the sectors share the same Minkowski vacuum completely fixed the free functions of time.
As a consequence, as in the two copies of GR, the general solution to the $\ell=0$ mode can be absorbed by a redefinition of the parameters in the Schwarzschild background, namely the two gravitational radii.
We also confirmed that the consequences of the linearized analysis for the $\ell=0$ mode can be naturally extended to the nonlinear case as the spherically-symmetric but time-dependent vacuum solutions written in the spatially-flat coordinates in the self-accelerating branch of MTBG.

In Sec.~\ref{sec5} and Sec.~\ref{sec6}, we have analyzed the even-parity perturbations for the $\ell=1$ and $\ell\geq 2$ modes, respectively.
Since the equations of motion for these modes have become quite involved, we have focused on the effectively massless case.
First, we have set the constant parameter $b$ which measures the ratio of the proper times between the two sectors to be unity.
By imposing that $b=1$, as in the case of the odd-parity perturbations \cite{Minamitsuji:2023lvi}, the effective mass term in the equations of motion for the perturbations vanishes.
However, even in the case of $b=1$, the terms depending on the Lagrange multipliers associated with the second-class constraints still remain nontrivial, and the essential structure of MTBG is maintained.
Second, we have also assumed that the background gravitational radii in the physical and fiducial sectors coincide.
With these assumptions, the system of the perturbed equations of motion in the even-parity sector has become somewhat tractable, while the essential features of the even-parity perturbations remain nontrivial.

In Sec.~\ref{sec5}, we have analyzed the even-parity perturbations for the $\ell=1$ mode in the effectively massless case.
In the effectively massless case we have exactly solved the set of equations of motion for the $\ell=1$ mode. 
However, the general solution for the $\ell=1$ mode contains several free functions of time.
Under the choice of the gauge \eqref{gauge_l1}, as a particular solution obtained by setting $\phi=0$ where the mode $\phi$ does not appear in the gauge-invariant combination $n_2-m_2$, these functions of time could be fixed by imposing the suitable boundary conditions at the spatial infinity, requiring that the leading order corrections to each component of the metrics are suppressed by a factor of $r^{-2}$ compared to the background quantities.
The gauge-invariant parts of the resultant solutions to the $\ell=1$ mode vanish, leaving no observable effect.

In Sec.~\ref{sec6}, we have analyzed the even-parity perturbations for the $\ell\geq 2$ modes. 
In the effectively massless case, in contrast to the cases of the $\ell=0$ and $\ell=1$ modes, we have found that there are two dynamical modes and four instantaneous modes.
We have also found the equations for the two dynamical modes $\Upsilon$ and $k_2$, and the elliptic equations for the four instantaneous modes, $\bar{\psi}$, $\bar{\chi}$, $\bar{k}_1$, and $\lambda_1$.
The equation for $\lambda_1$ is independent of other modes, and can be solved analytically.
We have shown that among the four solutions of $\lambda_1$, two of them are the solutions of the Laplace equation in the three-dimensional Euclid space and do not physically contribute to the dynamics of the other modes.
Since the number of the physically independent solutions of $\lambda_1$ is two, there is one instantaneous mode arising from $\lambda_1$.
Combined with the analysis of the odd-parity sector, in the effectively massless case there are four propagating DOFs and six instantaneous modes for each of the $\ell \geq 2$ modes in the self-accelerating branch of MTBG.

There are still many remaining issues.
The analysis of the even-parity perturbations should be extended by relaxing the assumptions of the effectively massless case.
The odd- and even-parity perturbations should be analyzed in the normal branch.
As an application of the even-party and odd-parity perturbations, BH quasinormal modes should be analyzed to distinguish BHs in MTBG from the case of the two copies of GR.
Finally, numerical simulation techniques for gravitational collapse in the self-accelerating and normal branch of MTBG should be developed.
These subjects would be definitively interesting but are left for future studies.

\section*{ACKNOWLEDGMENTS}
M.M.~was supported by the Portuguese national fund through the Funda\c{c}\~{a}o para a Ci\^encia e a Tecnologia in the scope of the framework of the Decree-Law 57/2016 of August 29, changed by Law 57/2017 of July 19, and the Centro de Astrof\'{\i}sica e Gravita\c c\~ao through the Project~No.~UIDB/00099/2020.
M.M.~also would like to acknowledge the Yukawa Institute for Theoretical Physics at Kyoto University for their hospitality, where a part of the present work has been performed during the long-term workshop `Gravity and Cosmology 2024.'
The work of S.M.~was supported in part by Japan Society for the Promotion of Science Grants-in-Aid for Scientific Research No.~24K07017 and the World Premier International Research Center Initiative (WPI), MEXT, Japan.


\bibliography{refs}

\begin{thebibliography}{23}%
\makeatletter
\providecommand \@ifxundefined [1]{%
 \@ifx{#1\undefined}
}%
\providecommand \@ifnum [1]{%
 \ifnum #1\expandafter \@firstoftwo
 \else \expandafter \@secondoftwo
 \fi
}%
\providecommand \@ifx [1]{%
 \ifx #1\expandafter \@firstoftwo
 \else \expandafter \@secondoftwo
 \fi
}%
\providecommand \natexlab [1]{#1}%
\providecommand \enquote  [1]{``#1''}%
\providecommand \bibnamefont  [1]{#1}%
\providecommand \bibfnamefont [1]{#1}%
\providecommand \citenamefont [1]{#1}%
\providecommand \href@noop [0]{\@secondoftwo}%
\providecommand \href [0]{\begingroup \@sanitize@url \@href}%
\providecommand \@href[1]{\@@startlink{#1}\@@href}%
\providecommand \@@href[1]{\endgroup#1\@@endlink}%
\providecommand \@sanitize@url [0]{\catcode `\\12\catcode `\$12\catcode
  `\&12\catcode `\#12\catcode `\^12\catcode `\_12\catcode `\%12\relax}%
\providecommand \@@startlink[1]{}%
\providecommand \@@endlink[0]{}%
\providecommand \url  [0]{\begingroup\@sanitize@url \@url }%
\providecommand \@url [1]{\endgroup\@href {#1}{\urlprefix }}%
\providecommand \urlprefix  [0]{URL }%
\providecommand \Eprint [0]{\href }%
\providecommand \doibase [0]{https://doi.org/}%
\providecommand \selectlanguage [0]{\@gobble}%
\providecommand \bibinfo  [0]{\@secondoftwo}%
\providecommand \bibfield  [0]{\@secondoftwo}%
\providecommand \translation [1]{[#1]}%
\providecommand \BibitemOpen [0]{}%
\providecommand \bibitemStop [0]{}%
\providecommand \bibitemNoStop [0]{.\EOS\space}%
\providecommand \EOS [0]{\spacefactor3000\relax}%
\providecommand \BibitemShut  [1]{\csname bibitem#1\endcsname}%
\let\auto@bib@innerbib\@empty
\bibitem [{\citenamefont {Fierz}\ and\ \citenamefont
  {Pauli}(1939)}]{Fierz:1939ix}%
  \BibitemOpen
  \bibfield  {author} {\bibinfo {author} {\bibfnamefont {M.}~\bibnamefont
  {Fierz}}\ and\ \bibinfo {author} {\bibfnamefont {W.}~\bibnamefont {Pauli}},\
  }\bibfield  {title} {\bibinfo {title} {{On relativistic wave equations for
  particles of arbitrary spin in an electromagnetic field}},\ }\href
  {https://doi.org/10.1098/rspa.1939.0140} {\bibfield  {journal} {\bibinfo
  {journal} {Proc. Roy. Soc. Lond. A}\ }\textbf {\bibinfo {volume} {173}},\
  \bibinfo {pages} {211} (\bibinfo {year} {1939})}\BibitemShut {NoStop}%
\bibitem [{\citenamefont {Boulware}\ and\ \citenamefont
  {Deser}(1972)}]{Boulware:1972yco}%
  \BibitemOpen
  \bibfield  {author} {\bibinfo {author} {\bibfnamefont {D.~G.}\ \bibnamefont
  {Boulware}}\ and\ \bibinfo {author} {\bibfnamefont {S.}~\bibnamefont
  {Deser}},\ }\bibfield  {title} {\bibinfo {title} {{Can gravitation have a
  finite range?}},\ }\href {https://doi.org/10.1103/PhysRevD.6.3368} {\bibfield
   {journal} {\bibinfo  {journal} {Phys. Rev. D}\ }\textbf {\bibinfo {volume}
  {6}},\ \bibinfo {pages} {3368} (\bibinfo {year} {1972})}\BibitemShut
  {NoStop}%
\bibitem [{\citenamefont {de~Rham}\ \emph {et~al.}(2011)\citenamefont
  {de~Rham}, \citenamefont {Gabadadze},\ and\ \citenamefont
  {Tolley}}]{deRham:2010kj}%
  \BibitemOpen
  \bibfield  {author} {\bibinfo {author} {\bibfnamefont {C.}~\bibnamefont
  {de~Rham}}, \bibinfo {author} {\bibfnamefont {G.}~\bibnamefont {Gabadadze}},\
  and\ \bibinfo {author} {\bibfnamefont {A.~J.}\ \bibnamefont {Tolley}},\
  }\bibfield  {title} {\bibinfo {title} {{Resummation of Massive Gravity}},\
  }\href {https://doi.org/10.1103/PhysRevLett.106.231101} {\bibfield  {journal}
  {\bibinfo  {journal} {Phys.Rev.Lett.}\ }\textbf {\bibinfo {volume} {106}},\
  \bibinfo {pages} {231101} (\bibinfo {year} {2011})},\ \Eprint
  {https://arxiv.org/abs/1011.1232} {arXiv:1011.1232 [hep-th]} \BibitemShut
  {NoStop}%
\bibitem [{\citenamefont {Hassan}\ and\ \citenamefont
  {Rosen}(2012)}]{Hassan:2011zd}%
  \BibitemOpen
  \bibfield  {author} {\bibinfo {author} {\bibfnamefont {S.}~\bibnamefont
  {Hassan}}\ and\ \bibinfo {author} {\bibfnamefont {R.}~\bibnamefont {Rosen}},\
  }\bibfield  {title} {\bibinfo {title} {Bimetric gravity from ghost-free
  massive gravity},\ }\href {https://doi.org/10.1007/JHEP02(2012)126}
  {\bibfield  {journal} {\bibinfo  {journal} {J. High Energy Phys.}\ }\textbf
  {\bibinfo {volume} {2012}}\bibfield  {number} {\bibinfo  {number} { (02)},\
  \bibinfo {eid} {126}},\ }\Eprint {https://arxiv.org/abs/1109.3515}
  {arXiv:1109.3515 [hep-th]} \BibitemShut {NoStop}%
\bibitem [{\citenamefont {Yamashita}\ \emph {et~al.}(2014)\citenamefont
  {Yamashita}, \citenamefont {De~Felice},\ and\ \citenamefont
  {Tanaka}}]{Yamashita:2014fga}%
  \BibitemOpen
  \bibfield  {author} {\bibinfo {author} {\bibfnamefont {Y.}~\bibnamefont
  {Yamashita}}, \bibinfo {author} {\bibfnamefont {A.}~\bibnamefont
  {De~Felice}},\ and\ \bibinfo {author} {\bibfnamefont {T.}~\bibnamefont
  {Tanaka}},\ }\bibfield  {title} {\bibinfo {title} {{Appearance of
  Boulware\textendash{}Deser ghost in bigravity with doubly coupled matter}},\
  }\href {https://doi.org/10.1142/S0218271814430032} {\bibfield  {journal}
  {\bibinfo  {journal} {Int. J. Mod. Phys. D}\ }\textbf {\bibinfo {volume}
  {23}},\ \bibinfo {pages} {1443003} (\bibinfo {year} {2014})},\ \Eprint
  {https://arxiv.org/abs/1408.0487} {arXiv:1408.0487 [hep-th]} \BibitemShut
  {NoStop}%
\bibitem [{\citenamefont {de~Rham}\ \emph {et~al.}(2014)\citenamefont
  {de~Rham}, \citenamefont {Heisenberg},\ and\ \citenamefont
  {Ribeiro}}]{deRham:2014fha}%
  \BibitemOpen
  \bibfield  {author} {\bibinfo {author} {\bibfnamefont {C.}~\bibnamefont
  {de~Rham}}, \bibinfo {author} {\bibfnamefont {L.}~\bibnamefont
  {Heisenberg}},\ and\ \bibinfo {author} {\bibfnamefont {R.~H.}\ \bibnamefont
  {Ribeiro}},\ }\bibfield  {title} {\bibinfo {title} {{Ghosts and matter
  couplings in massive gravity, bigravity and multigravity}},\ }\href
  {https://doi.org/10.1103/PhysRevD.90.124042} {\bibfield  {journal} {\bibinfo
  {journal} {Phys. Rev. D}\ }\textbf {\bibinfo {volume} {90}},\ \bibinfo
  {pages} {124042} (\bibinfo {year} {2014})},\ \Eprint
  {https://arxiv.org/abs/1409.3834} {arXiv:1409.3834 [hep-th]} \BibitemShut
  {NoStop}%
\bibitem [{\citenamefont {Gumrukcuoglu}\ \emph {et~al.}(2015)\citenamefont
  {Gumrukcuoglu}, \citenamefont {Heisenberg}, \citenamefont {Mukohyama},\ and\
  \citenamefont {Tanahashi}}]{Gumrukcuoglu:2015nua}%
  \BibitemOpen
  \bibfield  {author} {\bibinfo {author} {\bibfnamefont {A.~E.}\ \bibnamefont
  {Gumrukcuoglu}}, \bibinfo {author} {\bibfnamefont {L.}~\bibnamefont
  {Heisenberg}}, \bibinfo {author} {\bibfnamefont {S.}~\bibnamefont
  {Mukohyama}},\ and\ \bibinfo {author} {\bibfnamefont {N.}~\bibnamefont
  {Tanahashi}},\ }\bibfield  {title} {\bibinfo {title} {{Cosmology in bimetric
  theory with an effective composite coupling to matter}},\ }\href
  {https://doi.org/10.1088/1475-7516/2015/04/008} {\bibfield  {journal}
  {\bibinfo  {journal} {JCAP}\ }\textbf {\bibinfo {volume} {04}},\ \bibinfo
  {pages} {008}},\ \Eprint {https://arxiv.org/abs/1501.02790} {arXiv:1501.02790
  [hep-th]} \BibitemShut {NoStop}%
\bibitem [{\citenamefont {De~Felice}\ \emph
  {et~al.}(2021{\natexlab{a}})\citenamefont {De~Felice}, \citenamefont
  {Larrouturou}, \citenamefont {Mukohyama},\ and\ \citenamefont
  {Oliosi}}]{DeFelice:2020ecp}%
  \BibitemOpen
  \bibfield  {author} {\bibinfo {author} {\bibfnamefont {A.}~\bibnamefont
  {De~Felice}}, \bibinfo {author} {\bibfnamefont {F.}~\bibnamefont
  {Larrouturou}}, \bibinfo {author} {\bibfnamefont {S.}~\bibnamefont
  {Mukohyama}},\ and\ \bibinfo {author} {\bibfnamefont {M.}~\bibnamefont
  {Oliosi}},\ }\bibfield  {title} {\bibinfo {title} {{Minimal Theory of
  Bigravity: construction and cosmology}},\ }\href
  {https://doi.org/10.1088/1475-7516/2021/04/015} {\bibfield  {journal}
  {\bibinfo  {journal} {JCAP}\ }\textbf {\bibinfo {volume} {04}},\ \bibinfo
  {pages} {015}},\ \Eprint {https://arxiv.org/abs/2012.01073} {arXiv:2012.01073
  [gr-qc]} \BibitemShut {NoStop}%
\bibitem [{\citenamefont {Garcia-Saenz}\ \emph {et~al.}(2021)\citenamefont
  {Garcia-Saenz}, \citenamefont {Held},\ and\ \citenamefont
  {Zhang}}]{Garcia-Saenz:2021uyv}%
  \BibitemOpen
  \bibfield  {author} {\bibinfo {author} {\bibfnamefont {S.}~\bibnamefont
  {Garcia-Saenz}}, \bibinfo {author} {\bibfnamefont {A.}~\bibnamefont {Held}},\
  and\ \bibinfo {author} {\bibfnamefont {J.}~\bibnamefont {Zhang}},\ }\bibfield
   {title} {\bibinfo {title} {{Destabilization of Black Holes and Stars by
  Generalized Proca Fields}},\ }\href
  {https://doi.org/10.1103/PhysRevLett.127.131104} {\bibfield  {journal}
  {\bibinfo  {journal} {Phys. Rev. Lett.}\ }\textbf {\bibinfo {volume} {127}},\
  \bibinfo {pages} {131104} (\bibinfo {year} {2021})},\ \Eprint
  {https://arxiv.org/abs/2104.08049} {arXiv:2104.08049 [gr-qc]} \BibitemShut
  {NoStop}%
\bibitem [{\citenamefont {Silva}\ \emph {et~al.}(2022)\citenamefont {Silva},
  \citenamefont {Coates}, \citenamefont {Ramazano\u{g}lu},\ and\ \citenamefont
  {Sotiriou}}]{Silva:2021jya}%
  \BibitemOpen
  \bibfield  {author} {\bibinfo {author} {\bibfnamefont {H.~O.}\ \bibnamefont
  {Silva}}, \bibinfo {author} {\bibfnamefont {A.}~\bibnamefont {Coates}},
  \bibinfo {author} {\bibfnamefont {F.~M.}\ \bibnamefont {Ramazano\u{g}lu}},\
  and\ \bibinfo {author} {\bibfnamefont {T.~P.}\ \bibnamefont {Sotiriou}},\
  }\bibfield  {title} {\bibinfo {title} {{Ghost of vector fields in compact
  stars}},\ }\href {https://doi.org/10.1103/PhysRevD.105.024046} {\bibfield
  {journal} {\bibinfo  {journal} {Phys. Rev. D}\ }\textbf {\bibinfo {volume}
  {105}},\ \bibinfo {pages} {024046} (\bibinfo {year} {2022})},\ \Eprint
  {https://arxiv.org/abs/2110.04594} {arXiv:2110.04594 [gr-qc]} \BibitemShut
  {NoStop}%
\bibitem [{\citenamefont {Demirbo\u{g}a}\ \emph {et~al.}(2022)\citenamefont
  {Demirbo\u{g}a}, \citenamefont {Coates},\ and\ \citenamefont
  {Ramazano\u{g}lu}}]{Demirboga:2021nrc}%
  \BibitemOpen
  \bibfield  {author} {\bibinfo {author} {\bibfnamefont {E.~S.}\ \bibnamefont
  {Demirbo\u{g}a}}, \bibinfo {author} {\bibfnamefont {A.}~\bibnamefont
  {Coates}},\ and\ \bibinfo {author} {\bibfnamefont {F.~M.}\ \bibnamefont
  {Ramazano\u{g}lu}},\ }\bibfield  {title} {\bibinfo {title} {{Instability of
  vectorized stars}},\ }\href {https://doi.org/10.1103/PhysRevD.105.024057}
  {\bibfield  {journal} {\bibinfo  {journal} {Phys. Rev. D}\ }\textbf {\bibinfo
  {volume} {105}},\ \bibinfo {pages} {024057} (\bibinfo {year} {2022})},\
  \Eprint {https://arxiv.org/abs/2112.04269} {arXiv:2112.04269 [gr-qc]}
  \BibitemShut {NoStop}%
\bibitem [{\citenamefont {Manita}\ \emph {et~al.}(2022)\citenamefont {Manita},
  \citenamefont {Aoki}, \citenamefont {Fujita},\ and\ \citenamefont
  {Mukohyama}}]{Manita:2022tkl}%
  \BibitemOpen
  \bibfield  {author} {\bibinfo {author} {\bibfnamefont {Y.}~\bibnamefont
  {Manita}}, \bibinfo {author} {\bibfnamefont {K.}~\bibnamefont {Aoki}},
  \bibinfo {author} {\bibfnamefont {T.}~\bibnamefont {Fujita}},\ and\ \bibinfo
  {author} {\bibfnamefont {S.}~\bibnamefont {Mukohyama}},\ }\bibfield  {title}
  {\bibinfo {title} {{Spin-2 dark matter from anisotropic Universe in
  bigravity}},\ }\href@noop {} {\  (\bibinfo {year} {2022})},\ \Eprint
  {https://arxiv.org/abs/2211.15873} {arXiv:2211.15873 [gr-qc]} \BibitemShut
  {NoStop}%
\bibitem [{\citenamefont {De~Felice}\ \emph {et~al.}(2018)\citenamefont
  {De~Felice}, \citenamefont {Langlois}, \citenamefont {Mukohyama},
  \citenamefont {Noui},\ and\ \citenamefont {Wang}}]{DeFelice:2018ewo}%
  \BibitemOpen
  \bibfield  {author} {\bibinfo {author} {\bibfnamefont {A.}~\bibnamefont
  {De~Felice}}, \bibinfo {author} {\bibfnamefont {D.}~\bibnamefont {Langlois}},
  \bibinfo {author} {\bibfnamefont {S.}~\bibnamefont {Mukohyama}}, \bibinfo
  {author} {\bibfnamefont {K.}~\bibnamefont {Noui}},\ and\ \bibinfo {author}
  {\bibfnamefont {A.}~\bibnamefont {Wang}},\ }\bibfield  {title} {\bibinfo
  {title} {{Generalized instantaneous modes in higher-order scalar-tensor
  theories}},\ }\href {https://doi.org/10.1103/PhysRevD.98.084024} {\bibfield
  {journal} {\bibinfo  {journal} {Phys. Rev. D}\ }\textbf {\bibinfo {volume}
  {98}},\ \bibinfo {pages} {084024} (\bibinfo {year} {2018})},\ \Eprint
  {https://arxiv.org/abs/1803.06241} {arXiv:1803.06241 [hep-th]} \BibitemShut
  {NoStop}%
\bibitem [{\citenamefont {Minamitsuji}\ \emph {et~al.}(2023)\citenamefont
  {Minamitsuji}, \citenamefont {De~Felice}, \citenamefont {Mukohyama},\ and\
  \citenamefont {Oliosi}}]{Minamitsuji:2023lvi}%
  \BibitemOpen
  \bibfield  {author} {\bibinfo {author} {\bibfnamefont {M.}~\bibnamefont
  {Minamitsuji}}, \bibinfo {author} {\bibfnamefont {A.}~\bibnamefont
  {De~Felice}}, \bibinfo {author} {\bibfnamefont {S.}~\bibnamefont
  {Mukohyama}},\ and\ \bibinfo {author} {\bibfnamefont {M.}~\bibnamefont
  {Oliosi}},\ }\bibfield  {title} {\bibinfo {title} {{Gravitational collapse
  and odd-parity black hole perturbations in minimal theory of bigravity}},\
  }\href {https://doi.org/10.1103/PhysRevD.107.064070} {\bibfield  {journal}
  {\bibinfo  {journal} {Phys. Rev. D}\ }\textbf {\bibinfo {volume} {107}},\
  \bibinfo {pages} {064070} (\bibinfo {year} {2023})},\ \Eprint
  {https://arxiv.org/abs/2301.00498} {arXiv:2301.00498 [gr-qc]} \BibitemShut
  {NoStop}%
\bibitem [{\citenamefont {De~Felice}\ \emph {et~al.}(2020)\citenamefont
  {De~Felice}, \citenamefont {Doll},\ and\ \citenamefont
  {Mukohyama}}]{DeFelice:2020eju}%
  \BibitemOpen
  \bibfield  {author} {\bibinfo {author} {\bibfnamefont {A.}~\bibnamefont
  {De~Felice}}, \bibinfo {author} {\bibfnamefont {A.}~\bibnamefont {Doll}},\
  and\ \bibinfo {author} {\bibfnamefont {S.}~\bibnamefont {Mukohyama}},\
  }\bibfield  {title} {\bibinfo {title} {{A theory of type-II minimally
  modified gravity}},\ }\href {https://doi.org/10.1088/1475-7516/2020/09/034}
  {\bibfield  {journal} {\bibinfo  {journal} {JCAP}\ }\textbf {\bibinfo
  {volume} {09}},\ \bibinfo {pages} {034}},\ \Eprint
  {https://arxiv.org/abs/2004.12549} {arXiv:2004.12549 [gr-qc]} \BibitemShut
  {NoStop}%
\bibitem [{\citenamefont {De~Felice}\ \emph
  {et~al.}(2021{\natexlab{b}})\citenamefont {De~Felice}, \citenamefont
  {Mukohyama},\ and\ \citenamefont {Takahashi}}]{DeFelice:2021hps}%
  \BibitemOpen
  \bibfield  {author} {\bibinfo {author} {\bibfnamefont {A.}~\bibnamefont
  {De~Felice}}, \bibinfo {author} {\bibfnamefont {S.}~\bibnamefont
  {Mukohyama}},\ and\ \bibinfo {author} {\bibfnamefont {K.}~\bibnamefont
  {Takahashi}},\ }\bibfield  {title} {\bibinfo {title} {{Nonlinear definition
  of the shadowy mode in higher-order scalar-tensor theories}},\ }\href
  {https://doi.org/10.1088/1475-7516/2021/12/020} {\bibfield  {journal}
  {\bibinfo  {journal} {JCAP}\ }\textbf {\bibinfo {volume} {12}}\bibfield
  {number} {\bibinfo  {number} { (12)},\ \bibinfo {pages} {020}},\ }\Eprint
  {https://arxiv.org/abs/2110.03194} {arXiv:2110.03194 [gr-qc]} \BibitemShut
  {NoStop}%
\bibitem [{\citenamefont {Minamitsuji}\ \emph {et~al.}(2022)\citenamefont
  {Minamitsuji}, \citenamefont {De~Felice}, \citenamefont {Mukohyama},\ and\
  \citenamefont {Oliosi}}]{Minamitsuji:2022vfv}%
  \BibitemOpen
  \bibfield  {author} {\bibinfo {author} {\bibfnamefont {M.}~\bibnamefont
  {Minamitsuji}}, \bibinfo {author} {\bibfnamefont {A.}~\bibnamefont
  {De~Felice}}, \bibinfo {author} {\bibfnamefont {S.}~\bibnamefont
  {Mukohyama}},\ and\ \bibinfo {author} {\bibfnamefont {M.}~\bibnamefont
  {Oliosi}},\ }\bibfield  {title} {\bibinfo {title} {{Static and spherically
  symmetric general relativity solutions in minimal theory of bigravity}},\
  }\href {https://doi.org/10.1103/PhysRevD.105.123026} {\bibfield  {journal}
  {\bibinfo  {journal} {Phys. Rev. D}\ }\textbf {\bibinfo {volume} {105}},\
  \bibinfo {pages} {123026} (\bibinfo {year} {2022})},\ \Eprint
  {https://arxiv.org/abs/2204.08217} {arXiv:2204.08217 [gr-qc]} \BibitemShut
  {NoStop}%
\bibitem [{\citenamefont {De~Felice}\ \emph
  {et~al.}(2021{\natexlab{c}})\citenamefont {De~Felice}, \citenamefont
  {Mukohyama},\ and\ \citenamefont {Pookkillath}}]{DeFelice:2020cpt}%
  \BibitemOpen
  \bibfield  {author} {\bibinfo {author} {\bibfnamefont {A.}~\bibnamefont
  {De~Felice}}, \bibinfo {author} {\bibfnamefont {S.}~\bibnamefont
  {Mukohyama}},\ and\ \bibinfo {author} {\bibfnamefont {M.~C.}\ \bibnamefont
  {Pookkillath}},\ }\bibfield  {title} {\bibinfo {title} {{Addressing $H_0$
  tension by means of VCDM}},\ }\href
  {https://doi.org/10.1016/j.physletb.2021.136201} {\bibfield  {journal}
  {\bibinfo  {journal} {Phys. Lett. B}\ }\textbf {\bibinfo {volume} {816}},\
  \bibinfo {pages} {136201} (\bibinfo {year} {2021}{\natexlab{c}})},\ \Eprint
  {https://arxiv.org/abs/2009.08718} {arXiv:2009.08718 [astro-ph.CO]}
  \BibitemShut {NoStop}%
\bibitem [{\citenamefont {De~Felice}\ \emph
  {et~al.}(2021{\natexlab{d}})\citenamefont {De~Felice}, \citenamefont {Doll},
  \citenamefont {Larrouturou},\ and\ \citenamefont
  {Mukohyama}}]{DeFelice:2020onz}%
  \BibitemOpen
  \bibfield  {author} {\bibinfo {author} {\bibfnamefont {A.}~\bibnamefont
  {De~Felice}}, \bibinfo {author} {\bibfnamefont {A.}~\bibnamefont {Doll}},
  \bibinfo {author} {\bibfnamefont {F.}~\bibnamefont {Larrouturou}},\ and\
  \bibinfo {author} {\bibfnamefont {S.}~\bibnamefont {Mukohyama}},\ }\bibfield
  {title} {\bibinfo {title} {{Black holes in a type-II minimally modified
  gravity}},\ }\href {https://doi.org/10.1088/1475-7516/2021/03/004} {\bibfield
   {journal} {\bibinfo  {journal} {JCAP}\ }\textbf {\bibinfo {volume} {03}},\
  \bibinfo {pages} {004}},\ \Eprint {https://arxiv.org/abs/2010.13067}
  {arXiv:2010.13067 [gr-qc]} \BibitemShut {NoStop}%
\bibitem [{\citenamefont {De~Felice}\ \emph {et~al.}(2022)\citenamefont
  {De~Felice}, \citenamefont {Maeda}, \citenamefont {Mukohyama},\ and\
  \citenamefont {Pookkillath}}]{DeFelice:2022uxv}%
  \BibitemOpen
  \bibfield  {author} {\bibinfo {author} {\bibfnamefont {A.}~\bibnamefont
  {De~Felice}}, \bibinfo {author} {\bibfnamefont {K.-i.}\ \bibnamefont
  {Maeda}}, \bibinfo {author} {\bibfnamefont {S.}~\bibnamefont {Mukohyama}},\
  and\ \bibinfo {author} {\bibfnamefont {M.~C.}\ \bibnamefont {Pookkillath}},\
  }\bibfield  {title} {\bibinfo {title} {{Comparison of two theories of
  Type-IIa minimally modified gravity}},\ }\href
  {https://doi.org/10.1103/PhysRevD.106.024028} {\bibfield  {journal} {\bibinfo
   {journal} {Phys. Rev. D}\ }\textbf {\bibinfo {volume} {106}},\ \bibinfo
  {pages} {024028} (\bibinfo {year} {2022})},\ \Eprint
  {https://arxiv.org/abs/2204.08294} {arXiv:2204.08294 [gr-qc]} \BibitemShut
  {NoStop}%
\bibitem [{\citenamefont {Oppenheimer}\ and\ \citenamefont
  {Snyder}(1939)}]{Oppenheimer:1939ue}%
  \BibitemOpen
  \bibfield  {author} {\bibinfo {author} {\bibfnamefont {J.~R.}\ \bibnamefont
  {Oppenheimer}}\ and\ \bibinfo {author} {\bibfnamefont {H.}~\bibnamefont
  {Snyder}},\ }\bibfield  {title} {\bibinfo {title} {{On Continued
  gravitational contraction}},\ }\href {https://doi.org/10.1103/PhysRev.56.455}
  {\bibfield  {journal} {\bibinfo  {journal} {Phys. Rev.}\ }\textbf {\bibinfo
  {volume} {56}},\ \bibinfo {pages} {455} (\bibinfo {year} {1939})}\BibitemShut
  {NoStop}%
\bibitem [{\citenamefont {Kanai}\ \emph {et~al.}(2011)\citenamefont {Kanai},
  \citenamefont {Siino},\ and\ \citenamefont {Hosoya}}]{Kanai:2010ae}%
  \BibitemOpen
  \bibfield  {author} {\bibinfo {author} {\bibfnamefont {Y.}~\bibnamefont
  {Kanai}}, \bibinfo {author} {\bibfnamefont {M.}~\bibnamefont {Siino}},\ and\
  \bibinfo {author} {\bibfnamefont {A.}~\bibnamefont {Hosoya}},\ }\bibfield
  {title} {\bibinfo {title} {{Gravitational collapse in Painleve-Gullstrand
  coordinates}},\ }\href {https://doi.org/10.1143/PTP.125.1053} {\bibfield
  {journal} {\bibinfo  {journal} {Prog. Theor. Phys.}\ }\textbf {\bibinfo
  {volume} {125}},\ \bibinfo {pages} {1053} (\bibinfo {year} {2011})},\ \Eprint
  {https://arxiv.org/abs/1008.0470} {arXiv:1008.0470 [gr-qc]} \BibitemShut
  {NoStop}%
\bibitem [{\citenamefont {Blau}(2017)}]{Blau}%
  \BibitemOpen
  \bibfield  {author} {\bibinfo {author} {\bibfnamefont {M.}~\bibnamefont
  {Blau}},\ }\href@noop {} {\bibinfo {title} {Lecture notes on general
  relativity}} (\bibinfo {year} {2017})\BibitemShut {NoStop}%
\end{thebibliography}%
\end{document}